\documentclass[nohyperref]{article}

\usepackage{microtype}
\usepackage{graphicx}
\usepackage{subfigure}
\usepackage{booktabs} 
\usepackage{bm}
\usepackage{colortbl}
\usepackage{color}
\usepackage{bbm}
\usepackage{makecell}
\usepackage{hyperref}
\usepackage{multicol}

\makeatletter  
\newcommand\tabcaption{\def\@captype{table}\caption} 
\makeatother

\usepackage[accepted]{icml2022}

\usepackage{amsmath}
\usepackage{amssymb}
\usepackage{mathtools}
\usepackage{amsthm}

\usepackage[capitalize,noabbrev]{cleveref}

\theoremstyle{plain}

\theoremstyle{definition}

\theoremstyle{remark}

\usepackage[textsize=tiny]{todonotes}

\begin{document}

\twocolumn[
\icmltitle{Greedy-based Value Representation for Optimal Coordination \\ in Multi-agent Reinforcement Learning}



\icmlsetsymbol{equal}{*}

\begin{icmlauthorlist}
\icmlauthor{Lipeng Wan}{equal,xjtu}
\icmlauthor{Zeyang Liu}{equal,xjtu}
\icmlauthor{Xingyu Chen}{xjtu}
\icmlauthor{Xuguang Lan}{xjtu}
\icmlauthor{Nanning Zheng}{xjtu}
\end{icmlauthorlist}

\icmlaffiliation{xjtu}{School of Artificial Intelligence, Xian Jiaotong University, Xian, Shaanxi, China}

\icmlcorrespondingauthor{Xuguang Lan}{xglan@mail.xjtu.edu.cn}

\icmlkeywords{Multi-agent Reinforcement Learning, Credit Assignment}

\vskip 0.3in
]



\printAffiliationsAndNotice{\icmlEqualContribution} 

\begin{abstract}
    Due to the representation limitation of the joint Q value function, multi-agent reinforcement learning methods with linear value decomposition (LVD) or monotonic value decomposition (MVD) suffer from relative overgeneralization. As a result, they can not ensure optimal consistency (i.e., the correspondence between individual greedy actions and the maximal true Q value). In this paper, we derive the expression of the joint Q value function of LVD and MVD. According to the expression, we draw a transition diagram, where each self-transition node (STN) is a possible convergence. To ensure optimal consistency, the optimal node is required to be the unique STN. Therefore, we propose the greedy-based value representation (GVR), which turns the optimal node into an STN via inferior target shaping and further eliminates the non-optimal STNs via superior experience replay. In addition, GVR achieves an adaptive trade-off between optimality and stability. Our method outperforms state-of-the-art baselines in experiments on various benchmarks. Theoretical proofs and empirical results on matrix games demonstrate that GVR ensures optimal consistency under sufficient exploration.
\end{abstract}

\section{Introduction}
By taking advantage of the deep learning technique, cooperative multi-agent reinforcement learning (MARL) shows great scalability and excellent performance on challenging tasks \citep{robot1,traffic1} such as StarCraft unit micromanagement \citep{coma}. An essential problem of cooperative MARL is credit assignment. As a popular approach to address the problem, value decomposition gains growing attention. The main concern of value decomposition is the optimality of coordination. In a successful case of credit assignment via value decomposition, agents perform individual greedy actions according to their local utility functions and achieve the best team performance (i.e., the optimal true Q value). Here we define the correspondence between the individual greedy actions and the optimal true Q value as \textit{optimal consistency}.

Due to the representation limitation of the joint Q value function, linear value decomposition (LVD) or monotonic value decomposition (MVD) suffer from relative overgeneralization (RO) \citep{overgene1, overgene2}. As a result, they can not ensure the optimal consistency. Recent works address RO from two different perspectives. The first kind of method completes the representation capacity of the joint Q value function (e.g., QTRAN \citep{qtran} and QPLEX \citep{qplex}). However, learning the complete representation is impractical in complicated MARL tasks because the joint action space increases exponentially with the number of agents. The other kind of method tries to prevent sub-optimal convergences by learning a biased joint Q value function (e.g., WQMIX \citep{wqmix} and MAVEN \citep{maven}), which depends on heuristic parameters and is only applicable in specific tasks. More discussions about RO and related works are provided in Appendix A.

In this paper, we derive the expression of the joint Q value function of LVD and MVD and draw some interesting conclusions. Firstly, LVD and MVD share the same expression of the joint Q value function. Secondly, the joint Q value of any action depends on the true Q values of all actions in the whole joint action space. Thirdly, the joint Q values transfer with greedy actions, by which a transition diagram is acquired. In the transition diagram, each self-transition node (STN) is a possible convergence. A node that satisfies the optimal consistency is called an optimal node, otherwise, we call it a non-optimal node. To ensure the optimal consistency, the optimal node is required to be the unique STN, which is the \textit{target problem} of this paper. 

To address the target problem, we propose the greedy-based value representation (GVR). Firstly, we reshape the representation target of the inferior actions (i.e., the actions with poorer performance than current greedy), which is prove to ensure that the optimal node would always be an STN. 
We also prove that under the inferior target shaping, non-optimal nodes would be eliminated when the probabilities of superior actions (i.e., the actions with poorer performance than current greedy) exceed a threshold. Therefore, we introduce superior experience replay, which steadily raises the proportions of superior actions in the training batch. It is proved that GVR ensures the optimal consistency under sufficient exploration. However, excessive pursuit for optimality would weaken the stability in tasks with multiple optimums or multiple approximative optimums, for which we further design an adaptive trade-off between optimality and stability in GVR.

We have three contributions in this work. (1) This is the first work to derive the general expression of the joint Q value function for LVD and MVD. (2) Based on the expression of the joint Q value function, we draw a transition diagram and propose a quantified condition to ensure the optimal consistency for LVD and MVD. (3) We propose the GVR algorithm. GVR ensures the optimal consistency under sufficient exploration. Besides, GVR achieves an adaptive trade-off between optimality and stability. Our method outperforms state-of-the-art baselines in various benchmarks.

\section{Preliminaries}
\subsection{Dec-POMDP}
We model a fully cooperative multi-agent task as a decentralized partially observable Markov decision process (Dec-POMDP) described by a tuple $\mathcal{G} = <S, U, P, r, Z, O, n, \gamma>$ \citep{pomdp, decpomdp}. $s\in S$ denotes the true state of the environment. At each time step, each agent $a\in A\equiv \{1,2,\cdots,n\}$ receives a local observation $z^a \in Z$ produced by the observation function $O:S \times A \rightarrow Z$, and then chooses an individual action $u^a \in U$ according to a local policy $\pi^a(u^a|\tau^a): T\times U \rightarrow [0,1]$, where $\tau^a \in T\equiv (Z\times U)^*$ denotes the local action-observation history. The joint action of $n$ agents $\textbf{u}$ results in a shared reward $r(s,\textbf{u})$ and a transition to the next state $s' \sim P(\cdot|s, \textbf{u})$. $\gamma \in [0,1)$ is a discount factor.

We denote the joint variable of group agents with bold symbols, e.g., the joint action $\bm{u} \in \bm{U}\equiv U^n$, the joint action-observation history $\bm{\tau} \in \bm{\mathcal{T}}\equiv T^n$, and the joint policy (i.e., the policy interacts with environment to generate trajectories) $\bm{\pi}(\textbf{u}|\bm{\tau})$. The \textit{true Q value} is denoted by $\mathcal{Q}(s_t, \bm{u}_t)=\mathbb{E}_{s_{t+1:\infty},\bm{u}_{t+1:\infty}}\left[R_t|s_t,\bm{u}_t\right]$, where $R_t=\sum_{i=0}^\infty \gamma^i r_{t+1}$ is the discounted return. The action-state value function of agent $a$ and the group of agents are defined as \textit{utility function} $\mathcal{U}^a(u^a,\tau^a)$ and \textit{joint Q value function} $Q(\bm{u},\bm{\tau})$ respectively. The true Q value is the target of the joint Q value in training, which is the unique external criterion of the team performance. The \textit{greedy action} $\bm{\acute{u}}:=argmax_{\bm{u}}Q(\bm{u},\bm{\tau})$ is defined as the joint action with the maximal joint Q value . The \textit{optimal action} $\bm{u}^*:=argmax_{\bm{u}} \mathcal{Q}(s, \bm{u})$ is defined as the joint action with the best team performance. For brevity, we sometimes omit the prefix "joint" for the joint variables.

\subsection{Optimal Consistency and TGM Principle}
In centralized training with decentralized execution (CTDE) \citep{ctde1, ctde2, maddpg}, agents are expected to act individually according to their local policies (i.e., the individual greedy actions) while achieving the best team performance (i.e., the optimal true Q value). Here we define the correspondence between the individual greedy actions and the optimal true Q value as the optimal consistency.

\textbf{Definition 1 (Optimal consistency)}. Given a set of utility functions $\{\mathcal{U}^1(u^1,\tau^1)),\cdots,\mathcal{U}^n(u^n,\tau^n)\}$, and the true Q value $\mathcal{Q}(s,\bm{u})$, if the following holds
\begin{equation}
    \begin{aligned}
    &\{\underset{u^1}{argmax}\ \mathcal{U}^1(u^1,\tau^1),\cdots,\underset{u^n}{argmax}\ \mathcal{U}^n(u^n,\tau^n)\} \\
    & = \underset{\bm{u}}{argmax}\ \mathcal{Q}(s,\bm{u})\label{oc}
    \end{aligned}
\end{equation}
then we say the set of utility functions $\{\mathcal{U}^1(u^1,\tau^1)),\cdots,\mathcal{U}^n(u^n,\tau^n)\}$ satisfies the optimal consistency. For simplicity, we ignore situations with non-unique optimal actions.

The optimal consistency can be decomposed into two principles: Individual-Global-Max (IGM) and True-Global-Max (TGM). The IGM principle proposed by QTRAN \citep{qtran} is defined on the correspondence between individual greedy actions and the joint greedy actions (formally, $\{argmax_{u^1}\ \mathcal{U}^1(u^1,\tau^1),\cdots,argmax_{u^n}\ \mathcal{U}^n(o^n,\tau^n)\} = argmax_{\bm{u}} Q(\bm{u},\bm{\tau})$). To achieve the optimal consistency, the correspondence between the joint greedy action and the optimal true Q value is required, for which we define the TGM principle:

\textbf{Definition 2 (TGM)}. Given a joint value function $Q(\bm{u},\bm{\tau})$, and the true Q value $\mathcal{Q}(s,\bm{u})$, if the following holds
\begin{equation}
\underset{\bm{u}}{argmax}\ Q(\bm{u},\bm{\tau}) = \underset{\bm{u}}{argmax}\ \mathcal{Q}(s,\bm{u})\label{tgm}
\end{equation}
then we say the joint value function $Q(\bm{u},\bm{\tau})$ satisfies the TGM principle. For simplicity, we ignore situations with non-unique optimal actions.

\begin{figure*}[ht]
    \vskip 0.2in
    \begin{center}
    \centerline{\includegraphics[width=\columnwidth*5/3]{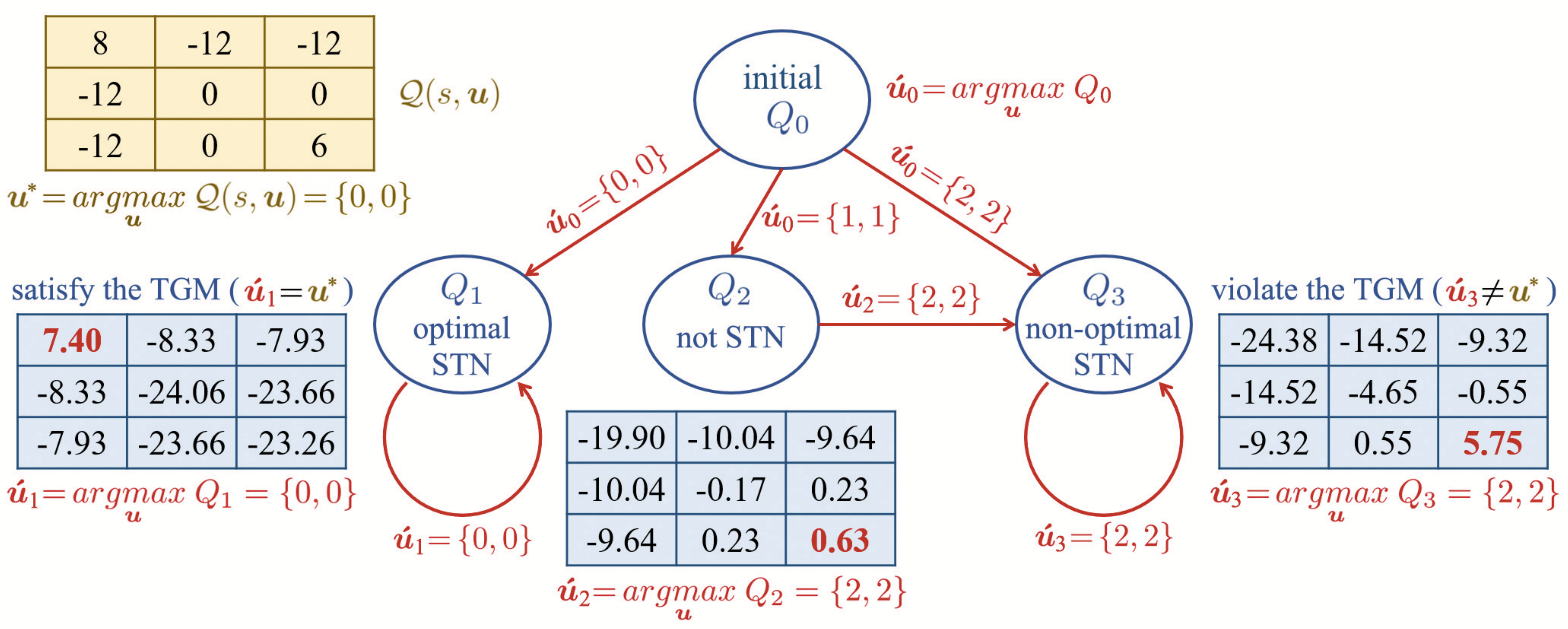}}
    \caption{Transition diagram of $Q(\bm{u},\bm{\tau})$ (blue tables, calculated by Eq.\ref{qfinal}). The true Q values $\mathcal{Q}(s,\bm{u})$ are shown in the yellow table. Initial $Q(\bm{u},\bm{\tau})=Q_0$ transfers to different nodes (i.e., $Q_1, Q_2, Q_3$) according to the corresponding greedy action $\bm{\acute{u}}_0=argmax_{\bm{u}}Q_0$. $Q_2$ is not a STN since it finally transfers to $Q_3$. $Q_1$ is the optimal STN while $Q_3$ is a non-optimal STN. We omit the other cases of $\bm{\acute{u}}_0$ such as $\bm{\acute{u}}_0=\{0,1\}$ and $\bm{\acute{u}}_0=\{2,0\}$.}
    \label{example}
    \end{center}
    \vskip -0.2in
\end{figure*}

\section{Investigation of the TGM Principle for LVD $\&$ MVD}
Linear value decomposition (LVD) and monotonic value decomposition (MVD) naturally meet the IGM principle \citep{qtran}. To achieve the optimal consistency, we investigate the conditions of the TGM principle for LVD and MVD. In this section, we first derive the expression of $Q(\bm{u},\bm{\tau})$ (Eq.\ref{qfinal}), by which we draw a transition diagram (Fig.\ref{example}) of the joint Q values. In this diagram, each self-transition node (STN) is a possible convergence, where the TGM principle is satisfied by the optimal node but violated by the non-optimal nodes. Finally, we propose a sufficient condition (Eq.\ref{remark}) of the TGM principle, which ensures the optimal node is the unique STN.

\subsection{Expression of the Joint Q Value Function for LVD $\&$ MVD}
Firstly, take two-agent LVD as an example, where the joint Q value function $Q(u^1_i,u^2_j,\bm{\tau})$ is linearly factorized into two utility functions as $Q(u^1_i,u^2_j,\bm{\tau})=\mathcal{U}^1(u^1_i,\tau^1)+\mathcal{U}^2(u^2_j,\tau^2)$. $u^1_i, u^2_j\in \{u_1,\cdots,u_m\}$ are the actions of agents 1 and 2, respectively, where $\{u_1,\cdots,u_m\}$ is the discrete individual action space. The greedy actions of two agents are denoted with $\acute{u}^1$ and $\acute{u}^2$. For briefness, we denote $\mathcal{Q}(s,u^1_i,u^2_j)$ and $\mathcal{U}^a(u^a_i,\tau^a)$ with $\mathcal{Q}_{ij}$ and $\mathcal{U}^a_i(a\in \{1,2 \})$, respectively.
Since the utility functions are trained to approximate different true Q values in different combinations, under $\epsilon-$greedy visitation, we have
\begin{equation}
\begin{aligned}
\mathcal{U}_i^1 = \frac{\epsilon}{m}\sum^m_{k=1}(\mathcal{Q}_{ik}-\mathcal{U}_k^2) + (1-\epsilon)(\mathcal{Q}_{i\acute{j}}-\mathcal{U}^2_{\acute{j}})\\
\mathcal{U}_j^2 = \frac{\epsilon}{m}\sum^m_{k=1}(\mathcal{Q}_{kj}-\mathcal{U}_k^1) + (1-\epsilon)(\mathcal{Q}_{\acute{i}j}-\mathcal{U}^1_{\acute{i}})\label{UaUb_main}
\end{aligned}
\end{equation} 
where the subscript $\acute{i}$ and $\acute{j}$ refer to the greedy actions of two agents. Through the derivation provided in Appendix B, the expression of $Q(u^1_i,u^2_j,\bm{\tau})$ can be acquired
\begin{equation}
\begin{aligned}
Q(u^1_i,u^2_j,\bm{\tau})=&\frac{\epsilon}{m}\sum_{k=1}^m(\mathcal{Q}_{ik}+\mathcal{Q}_{kj})+(1-\epsilon)(\mathcal{Q}_{\acute{i} j}+\mathcal{Q}_{i\acute{j}})\\
&-\frac{\epsilon(1-\epsilon)}{m}\sum_{k=1}^m(\mathcal{Q}_{\acute{i} k}+\mathcal{Q}_{k\acute{j}})\\
&-\frac{\epsilon^2}{m^2}\sum_{i=1}^m\sum_{j=1}^m \mathcal{Q}_{ij}-(1-\epsilon)^2\mathcal{Q}_{\acute{i}\acute{j}}\label{qfinal}
\end{aligned}
\end{equation}
Notice the term $\sum_{i=1}^m\sum_{j=1}^m \mathcal{Q}_{ij}$ depends on the true Q values of all actions in the whole joint action space. For MVD, the expression of $Q(u^1_i,u^2_j,\bm{\tau})$ is identical to Eq.\ref{qfinal} (the proof is provided in Appendix C.1). Verification of Eq.\ref{qfinal} is provided in Appendix C.2. For situations with more than two agents, by referring to the derivation in Appendix B and C.1, the expression of joint Q values can also be obtained.

\subsection{A Sufficient Condition of the TGM Principle for LVD $\&$ MVD}
According to Eq.\ref{qfinal}, the joint Q values $Q(\bm{u},\bm{\tau})(\bm{u}=\{u_i^1,u_j^2\})$ vary as the greedy action $\bm{\acute{u}}=\{\acute{u}^1,\acute{u}^2\}$, by which a transition diagram of $Q(\bm{u},\bm{\tau})$ is acquired. An example is shown in Fig.\ref{example}, where $\epsilon=0.2$.

\begin{figure}[t]
    \vskip 0.2in
    \begin{center}
    \centerline{\includegraphics[width=\columnwidth]{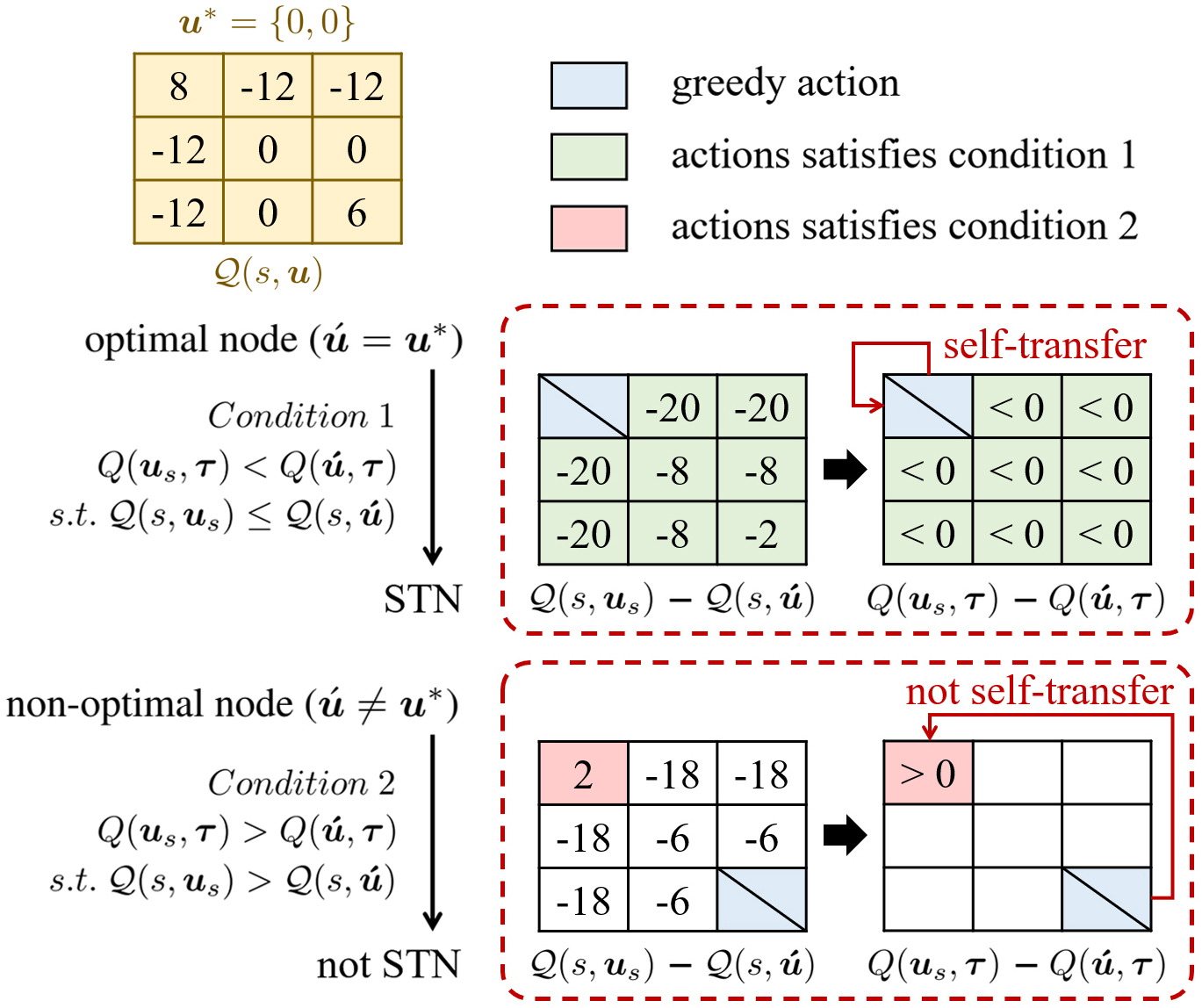}}
    \caption{Examples of joint Q value functions which satisfy $Condition\ 1$ (upper dotted box) and $Condition\ 2$ (lower dotted box), where $\bm{u}_s\neq\bm{\acute{u}}$. The true Q values are shown in the yellow table.}\label{example3}
    \end{center}
    \vskip -0.2in
\end{figure}

Notice $Q_1$ and $Q_3$ in Fig.\ref{example} are both STNs. An STN satisfies 
\begin{equation}
\underset{\bm{u}}{argmax}\ Q(\bm{u},\bm{\tau}) = \underset{\bm{u}}{argmax}\ Q_{old}(\bm{u},\bm{\tau})\label{de3}
\end{equation}
where $Q_{old}(\bm{u},\bm{\tau})$ and $Q(\bm{u},\bm{\tau})$ are the joint Q value functions of the nodes before and after transition, respectively. Especially, if $Q(\bm{u},\bm{\tau})$ is an STN and satisfies the TGM principle (i.e., $\bm{\acute{u}}=\bm{u}^*$), we say it is the optimal STN. Otherwise, if $Q(\bm{u},\bm{\tau})$ is an STN but violates the TGM principle (i.e., $\bm{\acute{u}}\neq\bm{u}^*$), we say it is a non-optimal STN. Each STN is a possible convergence. To ensure the TGM principle, \textit{the optimal node is required to be the unique STN}, i.e., the optimal node must be the only possible convergence. To achieve this, we provide a set of sufficient conditions as follows: for $\forall \bm{u}_s\neq\bm{\acute{u}}$,
\begin{equation}
    \begin{aligned}
    &\#\ Condition\ 1 \\
    &Q(\bm{u}_s,\bm{\tau})<Q(\bm{\acute{u}},\bm{\tau})\ \ \ \ \ \ \ \ \ s.t.\ \mathcal{Q}(s,\bm{u}_s)\leq\mathcal{Q}(s,\bm{\acute{u}}) \\
    &\#\ Condition\ 2 \\
    &Q(\bm{u}_s,\bm{\tau})>Q(\bm{\acute{u}},\bm{\tau})  \ \ \ \ \ \ \ \ \ s.t.\ \mathcal{Q}(s,\bm{u}_s)>\mathcal{Q}(s,\bm{\acute{u}}) 
    \end{aligned}
\label{remark}
\end{equation}
Examples which satisfy $Condition\ 1$ and $Condition\ 2$ are shown in Fig.\ref{example3}.

$\bm{Condition\ 1}$ ensures the optimal node is an STN. Suppose $\bm{\acute{u}}=argmax_{\bm{u}}Q_{old}(\bm{u},\bm{\tau})$ equals to the optimal action, i.e., $\bm{\acute{u}}=\bm{u}^*$, we have $\forall \bm{u}_s\neq\bm{\acute{u}}$ ,$\mathcal{Q}(s,\bm{u}_s)<\mathcal{Q}(s,\bm{\acute{u}})$. According to $Condition\ 1$, $\forall \bm{u}_s \neq \bm{u}^*$, $Q(\bm{u}_s,\bm{\tau})<Q(\bm{\acute{u}},\bm{\tau})$. By Eq.\ref{de3}, the optimal node is an STN since $argmax_{\bm{u}}Q(\bm{u},\bm{\tau})=\bm{u}^*=argmax_{\bm{u}}Q_{old}(\bm{u},\bm{\tau})$.

$\bm{Condition\ 2}$ ensures a non-optimal node is not STN. Suppose $\bm{\acute{u}}=argmax_{\bm{u}}Q_{old}(\bm{u},\bm{\tau})$ is a non-optimal action, $\exists \bm{u}_s$ such that $\mathcal{Q}(s,\bm{u}_s)>\mathcal{Q}(s,\bm{\acute{u}})$. According to $Condition\ 2$, $Q(\bm{u}_s,\bm{\tau})>Q(\bm{\acute{u}},\bm{\tau})$. By Eq.\ref{de3}, the non-optimal node is not an STN since $argmax_{\bm{u}}Q(\bm{u},\bm{\tau})\neq argmax_{\bm{u}}Q_{old}(\bm{u},\bm{\tau})$. 

\section{Greedy-based Value Representation}
It is impractical to evaluate the conditions in Eq.\ref{remark} for LVD or MVD because both $Q(\bm{u}_s,\bm{\tau})$ and $Q(\bm{\acute{u}},\bm{\tau})$ depend on the true Q values of all actions in the whole joint action space. In this section, we first introduce inferior target shaping (ITS), where $Q(\bm{u}_s,\bm{\tau})-Q(\bm{\acute{u}},\bm{\tau})$ becomes independent to the true Q values of inferior actions. We prove that $Condition\ 1$ always holds under ITS and $Condition\ 2$ can be further satisfied by superior experience replay (SER). Besides, as discussed in Appendix H.2, excessive pursuit for optimality decreases the stability. In Section 4.3,  we introduce a method to achieve an adaptive trade-off between stability and optimality in GVR. An overview of GVR is given in Appendix I.

\subsection{Inferior Target Shaping}
According to Eq.\ref{qfinal}, the joint Q values of the optimal node depend on the true Q values of all actions in the whole joint action space. We can turn the optimal node into an STN by modifying some of these true Q values. Since the exact true Q values of non-optimal actions are uninformative, we reshape them with an ITS target $\mathcal{Q}_{its}(s, \bm{u})$
\begin{equation}
\begin{aligned}
&\mathcal{Q}_{its}(s, \bm{u})=Q(\bm{\acute{u}},\bm{\tau})-\alpha |Q(\bm{\acute{u}},\bm{\tau})| \\
&\ \ \ \ s.t.\ \ \mathcal{Q}(s,\bm{u})\leq \mathcal{Q}(s, \bm{\acute{u}})*(1+e_{Q0})\ and \ \bm{u} \neq \bm{\acute{u}}
\end{aligned}
\label{its}
\end{equation}
$\mathcal{Q}(s,\bm{u})$ and $\mathcal{Q}(s, \bm{\acute{u}})$ can be acquired by estimation, which is discussed in Section 4.3. An action satisfying the constraints in Eq.\ref{its} is called an \textit{inferior action}. Similarly, an action $\bm{u}$ satisfying $\mathcal{Q}(s, \bm{u})>\mathcal{Q}(s,\bm{\acute{u}})*(1+e_{Q0})$ and $\bm{u}\neq\bm{\acute{u}}$ is called \textit{superior action}. For greedy or superior actions, $\mathcal{Q}_{its}(s,\bm{u})=\mathcal{Q}(s, \bm{u})$. $\alpha\in(0,\infty)$ is a hyper-parameter that defines the gap of the joint Q values' targets between the inferior and greedy actions. $e_{Q0}\in[0,\infty)$ is a hyper-parameter that defines the minimum gap of the true Q values between the superior and greedy actions. An example of the ITS target is provided in Appendix D.1. The ITS target simplifies the representation since there is no need to represent the true Q values of inferior actions. Besides, given the greedy action $\bm{\acute{u}}$ and another action $\bm{u}_s (\bm{u}_s\neq \bm{\acute{u}})$, assuming $\mathcal{Q}(s, \bm{\acute{u}})>0$, we have
\begin{equation}
\begin{aligned}
Q(\bm{u}_s,&\bm{\tau})-Q(\bm{\acute{u}},\bm{\tau})\\
=&n(\eta_1-\eta_2)\left[\mathcal{Q}(s, \bm{\acute{u}})-(1-\alpha)Q(\bm{\acute{u}},\bm{\tau})\right]\\ &+ n\eta_1 e_Q\mathcal{Q}(s,\bm{\acute{u}})\label{con_nagents}
\end{aligned}
\end{equation}
where $e_Q=\frac{\mathcal{Q}_{its}(s, \bm{u}_s)-\mathcal{Q}(s, \bm{\acute{u}})}{\mathcal{Q}(s, \bm{\acute{u}})}$,$\eta_1=(\frac{\epsilon}{m})^{n-1}$, and $\eta_2=(1-\epsilon+\frac{\epsilon}{m})^{n-1}$. We provide two different derivations for Eq.\ref{con_nagents} in Appendix D, and the calculation result is verified in the experimental part (Fig.\ref{exp1.3}(a)). 

Notice $Q(\bm{u}_s,\bm{\tau})-Q(\bm{\acute{u}},\bm{\tau})$ is \textit{independent} to the true Q values of inferior actions. It is proved in Appendix E.1 that $Condition\ 1$ (Eq.\ref{remark}) always holds under ITS, i.e., the optimal node is \textit{always an STN}. Besides, we also prove that $Condition\ 2$ holds when 
\begin{equation}
\frac{\eta_1}{\eta_2}>\frac{\alpha}{\alpha+e_{Q0}}=\eta_0\label{e_00s}
\end{equation}
which indicates the non-optimal STNs can be eliminated by \textit{raising $\frac{\eta_1}{\eta_2}$ under ITS}. A simple way to achieve this is improving exploration. Substituting the expression of $\eta_1$ and $\eta_2$ into Eq.\ref{e_00s}, we have
\begin{equation}
\epsilon>\frac{m}{({\frac{e_{Q0}}{\alpha})^\frac{1}{n-1}+1}+m-1}=\epsilon_0\label{e_0}
\end{equation}
where $\epsilon_0$ is the lower bound of $\epsilon$. However, as the number of agents ($n$) and the size of individual action space ($m$) increases, $\epsilon_0$ grows close to 1 (an example is provided in Fig.\ref{exp1.3}(b)), which suggests improving exploration is inapplicable in tasks with long episodes. 

Another way to raise $\frac{\eta_1}{\eta_2}$ is applying a weight $w(w>1)$ to the superior actions. It is proved in Appendix F.1 that the non-optimal STNs can be eliminated when $w>\frac{\alpha(\eta_2-\eta_1)}{e_Q\eta_1}=w_0$, where $w_0$ is the lower bound of $w$. However, $w_0$ grows exponentially as the number of agents increases (as verified in Appendix F.2, $w_0=659.50$ when $n=4$), which introduces instability in $Q(\bm{u},\bm{\tau})$.

\subsection{Superior Experience Replay}
The difficulty of meeting Eq.\ref{e_00s} is that $\eta_0$ is a constant while $\frac{\eta_1}{\eta_2}$ decreases exponentially with the number of agents. We consider adding a constant to $\eta_1$, i.e., adding a constant to the probabilities of the superior actions. Therefore, we introduce an extra buffer, named \textit{superior buffer}, to store the superior actions. Inspired by prioritized experience replay (PER) \citep{per, erp}, we assign a priority to each trajectory. The training batch consists of two parts: trajectories randomly sampled from the replay buffer and trajectories sampled from the superior buffer with PER.

We apply a weight $w_{ser}$ to the samples from the superior buffer. The probability of sampling the superior actions from the replay buffer is extremely small. Therefore, the proportion of the superior actions in the training batch is mainly determined by $w_{ser}$, which is a constant. It is proved in Appendix G that under ITS, SER can eliminate the non-optimal STNs by setting $w_{ser}$ to
\begin{equation}
w_{ser}>\frac{\alpha}{e_{Q0}}(\eta_2-\eta_1)\eta_s-\eta_1\eta_s\label{conditionser}
\end{equation}
where $\eta_s$ is the probability of state $s$. 

\subsection{Adaptive Trade-off between Optimality and Stability}
In Eq.\ref{its}, both $\mathcal{Q}(s,\bm{u})$ and $\mathcal{Q}(s, \bm{\acute{u}})$ are unavailable, which require estimation. Excessive pursuit for optimality decreases the stability due to the error of the estimation. An example is provided in Appendix H.2, where the joint Q values frequently transfer among the reckoned "optimal" nodes and poor intermediate nodes. As a result, methods without a trade-off between optimality and stability may suffer from poor performance.

In Eq.\ref{its}, notice $e_{Q0}$ defines the minimum gap of the true Q values between the superior and greedy actions. When the estimation error of the true Q values exceeds $e_{Q0}$, an inferior action may be mistaken for a superior action, which causes instability. As $e_{Q0}$ grows, the actions just slightly better than current greedy action are classified into inferior actions, which improves the stability but decreases the optimality. 
To approximate the optimality on the premise of stability, we introduce an uncertainty-based trade-off to learn an adaptive $e_{Q0}$.

\begin{figure}[t]
    \vskip 0.2in
    \begin{center}
    \centerline{\includegraphics[width=\columnwidth]{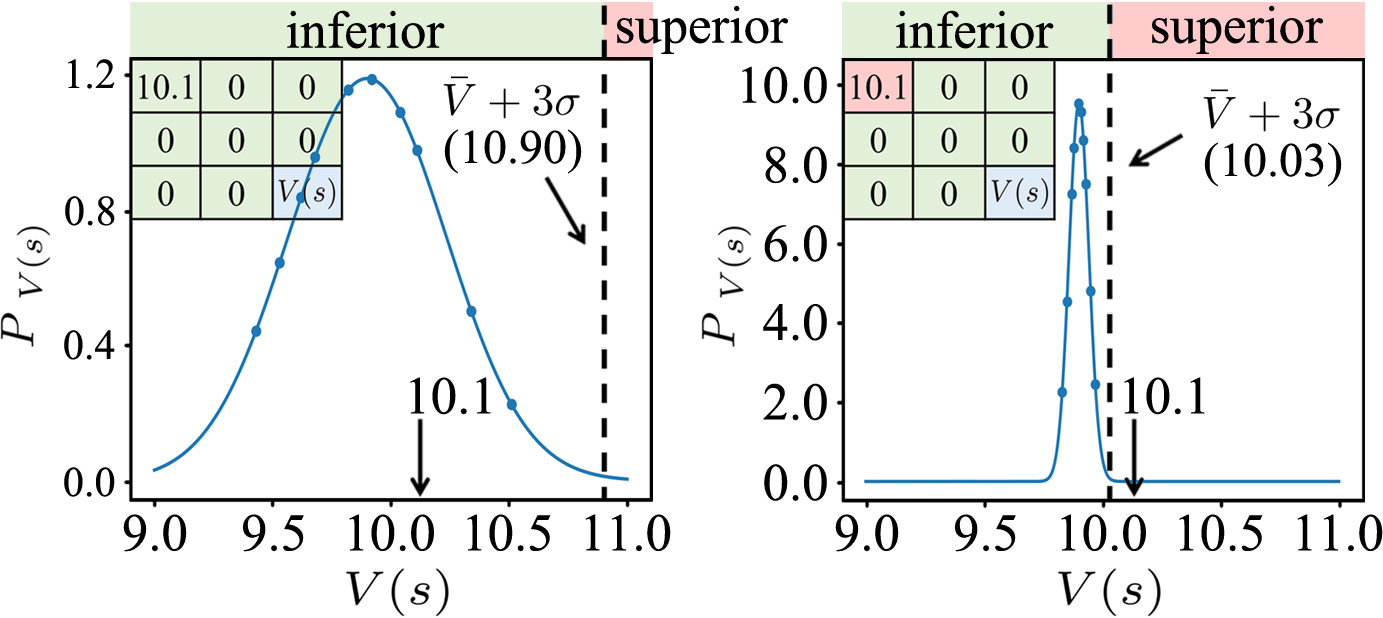}}
    \caption{An example of the trade-off between optimality and stability in GVR. According to the outputs of the critics (blue dots), a gaussian distribution (blue curve) is fitted, where the threshold of superior and inferior actions is set to $\bar{V}+3\sigma$. When $Q(s,\bm{u}_s)>\bar{V}+3\sigma$, the probability of $u_s$ being a superior action is 0.9987.}
    \label{tradeoff}
    \end{center}
    \vskip -0.2in
\end{figure}

Inspired by ensemble learning, we apply a group of centralized critics $\{V_1, V_2,\cdots,V_{n_c}\}$ to estimate $\mathcal{Q}(s, \bm{\acute{u}})$, where $n_c$ is the number of the critics. The target of $V_c(s_{t_0})$ ($c\in[1,n_c]$) is
\begin{equation}
\mathcal{V}_{tar}(s_{t_0})= 
\left\{
\begin{array}{lr}
\sum_{t=t_0}^T \gamma^{t-t_0}r(s_t,\bm{u}_t) \ \ \ \ \bm{u} = \bm{\acute{u}}\\ 
V_c(s_{t_0})\ \ \ \ \ \ \ \ \ \ \ \ \ \ \ \ \ \ \ \ \ \ \ \ \ \ \ \bm{u} \neq \bm{\acute{u}}
\end{array}
\right.	\label{vtarget}
\end{equation} 
Assume $\{V_1(s_t), V_2(s_t),\cdots,V_{n_c}(s_t)\}$ obey the Gaussian distribution $\mathcal{N}(\bar{V}(s_t), \sigma(s_t))$, where the mean $\bar{V}(s_t)$ and variance $\sigma(s_t)$ can be acquired by maximum likelihood estimation. Instability occurs when an inferior action is misclassified into a superior action. The uncertainty of misclassification can be represented by $\sigma(s_t)$. According to the $3$-$\sigma$ rule, we define $e_{Q0}(s_t)$ as
\begin{equation}
e_{Q0}(s_t) = \frac{3\sigma(s_t)}{\bar{V}(s_t)}\label{eq0}
\end{equation}
According to Eq.\ref{its}, an action $\bm{u}_t$ is classified as the superior action when $\sum_{t=t_0}^T \gamma^{t-t_0} r(s_t,\bm{u}_t)>\bar{V}(s_t)+3\sigma(s_t)$. Since $P(\bar{V}(s_t)-3 \sigma(s_t) <V(s_t)<\bar{V}(s_t)+3\sigma(s_t))=0.997$, the theoretical probability of misclassification (without considerings the estimation bias of the critics) equals $P(V(s_t)>\bar{V}(s_t)+3\sigma(s_t))=\frac{1-0.997}{2}=0.135\%$. As a result, an action will not be accepted as a superior action without sufficient confidence. The critics are more cautious when faced with a unfamiliar state, which prevents instability adaptively. An example is shown in Fig.\ref{tradeoff}. We provide investigations of many other threshold functions in appendix J.2. 

The algorithm is given in Algo.\ref{algorithm}, where $\mathbbm{I}_{sup}(s_t,\bm{u}_t)$ is a indicator for the superior action, i.e., $\mathbbm{I}_{sup}(s_t,\bm{u}_t)=1$, $s.t. \sum_{t=t_0}^T \gamma^{t-t_0} r(s_t,\bm{u}_t)>\bar{V}(s_t)+3\sigma(s_t)$. $w_{ser}(s_t)$ and $e_{Q0}(s_t)$ are calculated by Eq.\ref{conditionser} and Eq.\ref{eq0} respectively, where $\eta_s=1$. The priority of trajectory $\bm{\tau}$ equals to $p_{\bm{\tau}}=\sum_t^{\bm{\tau}}\mathbbm{I}_{sup}(s_t,\bm{u}_t)\left[\sum_{t=t_0}^T \gamma^{t-t_0} r(s_t,\bm{u}_t)-\bar{V}(s_t)-3\sigma(s_t)\right]$. For brevity, we omit the inputs of $\mathcal{V}_{tar}$, $\mathcal{Q}_{its}$ and $\mathbbm{I}_{sup}$.

\begin{algorithm}[t]
\caption{Greedy-based Value Representation}
\label{algorithm}
\begin{algorithmic}
\STATE {Initialize parameters $\theta_a$ and $\{\phi_1,\cdots,\phi_{c_n}\}$}
\STATE {Initialize replay buffer $D_r$ and superior buffer $D_s$}
\FOR{Iterations $i = 1,2,\cdots$}
	\IF {test rounds}
		\FOR{critic $c = 1,2, \cdots , n_c$}
			\STATE{Sample batch $b_c$ from test trajectories $\bm{\tau}_{test}$}
			\STATE{$loss_c = \sum_{\bm{\tau}}^{b_c}\sum_t^{\bm{\tau}}\left[\mathcal{V}_{tar}-V_{\phi_c}(s_t)\right]$}
		\ENDFOR
		\ELSE
		\STATE{Sample batch $b_r$ randomly from $D_r$}
		\STATE{$loss_{a1} =  \sum_{\bm{\tau}}^{b_r} \sum_t^{\bm{\tau}} \left[ \mathcal{Q}_{its}-Q_{\theta_a}({\bm{u}_t,\bm{\tau}_t})\right]$}
		\STATE{Take out $\bm{\tau}_s$ with the top priority from $D_s$}
		\STATE{$loss_{a2} = \sum_t^{\bm{\tau}_s} w_{ser}(s_t) \mathbbm{I}_{sup}\left[Q_{its}- Q_{\theta_a}(\bm{u}_t,\bm{\tau}_t) \right]$}
		\STATE{$loss_a = loss_{a1} + loss_{a2}$}
		\STATE {Calculate priorities for \{$b_r$, $\tau_s$\} and store it to $D_s$}
	\ENDIF
	\STATE {Update $\theta_a$ and $\{\phi_1,\cdots,\phi_{c_n}\}$}
	\ENDFOR
\end{algorithmic}
\end{algorithm}

\section{Experiments}
In this section, firstly we verify our conclusion about the STNs in matrix games under difference conditions, where we also evaluate the effectiveness of GVR. Secondly, to evaluate the stability and scalability of our method, we test the performance of GVR in predator-prey tasks with extreme reward shaping and challenging tasks of StarCraft multi-agent challenge (SMAC) \citep{smac}. Finally, we design ablation studies to investigate the improvements of GVR. Our method is compared with state-of-the-art baselines including QMIX \citep{qmix}, QPLEX \citep{qplex}, and WQMIX \citep{wqmix}. All results are evaluated over 5 seeds. Experimental settings and more experiments are provided in Appendix J.

\begin{figure}[t]
    \vskip 0.2in
    \begin{center}
    \centerline{\includegraphics[width=\columnwidth]{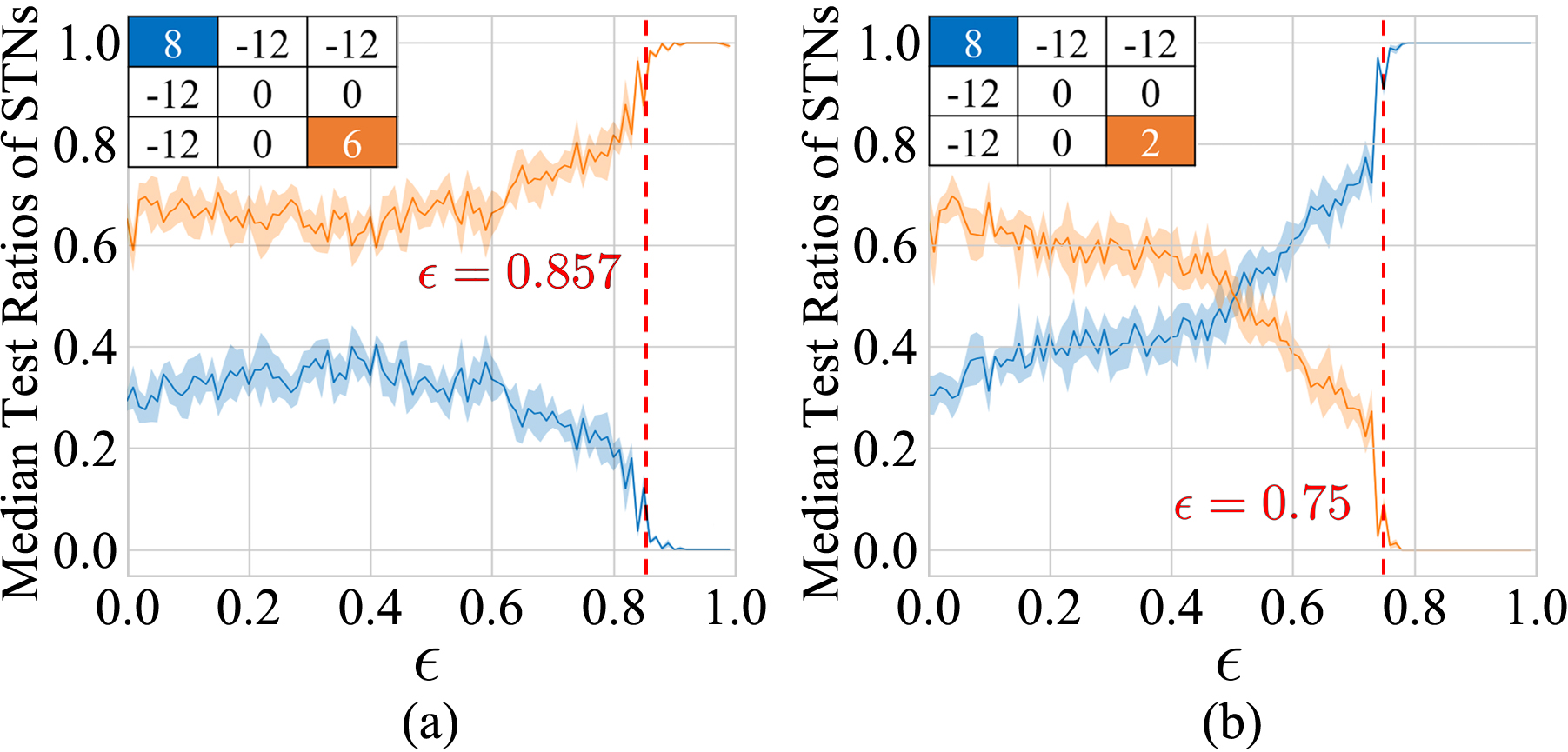}}
    \caption{Median test ratios of the STNs. The pay-off matrices are shown in the upper-left of sub-graphs. In sub-graph (a), when $\epsilon=0.4$, the ratios of the optimal STNs (blue) and the non-optimal STN (orange) approximate 0.35 and 0.65 respectively. There remains \textit{only one} STN when $\epsilon>0.857$, which consists with the calculated threshold (denoted by red dotted lines) from Eq.\ref{qfinal}.}\label{eplow}
    \end{center}
    \vskip -0.2in
\end{figure}

\subsection{One-step Matrix Game}
Matrix game is a simple fully cooperative multi-agent task, where the shared rewards are defined by a pay-off matrix. In one-step matrix games, the true $Q$ values are identical to the shared reward, which is convenient for the verification of the optimal consistency. We denote the size of a matrix by $m^n$, where $n$ is the number of agents and $m$ is the size of individual action space.

\begin{figure}[h]
    \vskip 0.2in
    \begin{center}
    \centerline{\includegraphics[width=\columnwidth]{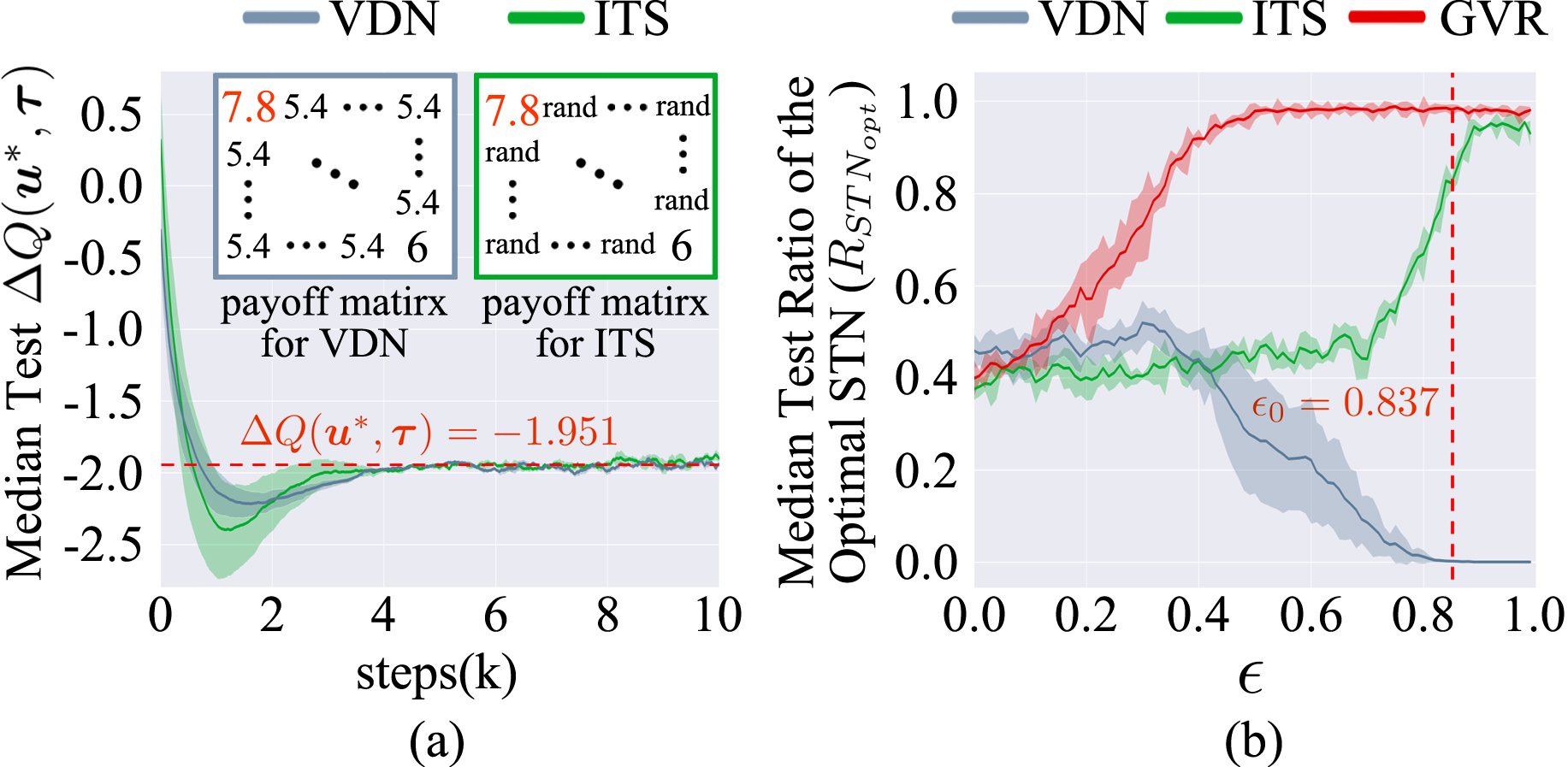}}
    \caption{Evaluation of GVR in 4-agent one-step matrix games. (a) Test $\Delta Q(\bm{u}^*,\bm{\tau})$ under ITS with random rewards. (b) Median test ratio of the optimal STN ($R_{STN_{opt}}$) as $\epsilon$ grows. $R_{STN_{opt}}=0$ denotes the optimal node is not an STN while $R_{STN_{opt}}=1$ denotes the optimal STN is the unique STN.}
    \label{exp1.3}
    \end{center}
    \vskip -0.2in
\end{figure}

\begin{figure*}[t]
    \vskip 0.2in
    \begin{center}
    \centerline{\includegraphics[width=\columnwidth*20/13]{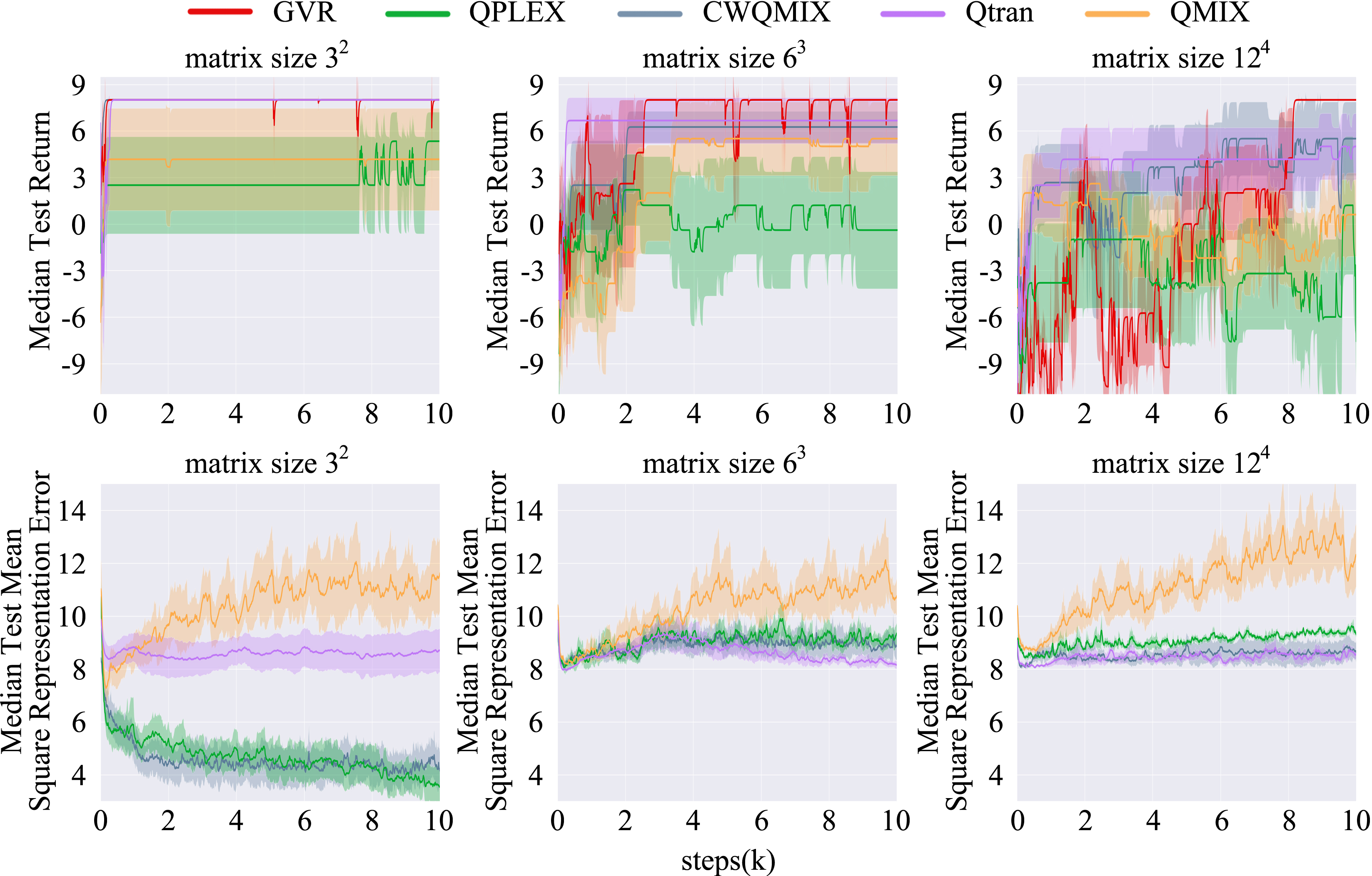}}
    \caption{GVR vs methods with complete representation capacity in matrix games of size $3^2$, $6^3$, and $12^4$.}
    \label{exp1.4}
    \end{center}
    \vskip -0.2in
\end{figure*}

\textbf{The verification of STNs under difference conditions.}
We conduct experiments on two-agent one-step matrix game to verify the expression of joint Q values (i.e., Eq.\ref{qfinal}) of LVD and MVD. The experimental results are provided in Appendix C.2. Besides, we test ratios of different STNs (i.e., the probabilities of different convergence) for LVD as $\epsilon$ increasing from 0.01 to 1. As shown in Fig.\ref{eplow}, some of the STNs disappear under large exploration, and there remains only one STN when $\epsilon$ approaches 1. However, \textit{large exploration do not ensure the optimal consistency} because the remained STN is not necessarily the optimal STN.

\textbf{Evaluation of GVR.}
We first verify the expression of $Q(\bm{u}_s,\bm{\tau})-Q(\bm{\acute{u}},\bm{\tau})$ (Eq.\ref{con_nagents}). Two pay-off matrices of size $3^4$ are generated for VDN and ITS according to Tab.\ref{rand_mat}, where $\bm{u}_s =\bm{u}^*=\{0,0,0,0\}$, $\bm{\acute{u}}=\{2,2,2,2\}$, $e_Q=0.3$ and $\alpha=0.1$. We measure $\Delta Q(\bm{u}^*,\bm{\tau}):=Q(\bm{u}^*,\bm{\tau})-Q(\bm{\acute{u}},\bm{\tau})$ for VDN and ITS trained with corresponding matrices. As shown in Fig.\ref{exp1.3}(a), the test results of $\Delta Q(\bm{u}^*,\bm{\tau})$ consists with the calculation result from Eq.\ref{con_nagents} ($\Delta Q(\bm{u}^*,\bm{\tau})=-1.951$, denoted by the red dotted line). Besides, the test results are the same under different random seeds for ITS, which indicates the conditions in Eq.\ref{remark} are independent to the true Q values of inferior actions for ITS.

\begin{table}[h]
\caption{Pay-off matrices generated for LVD and ITS.}
\label{rand_mat}
\vskip 0.15in
\begin{center}
\begin{small}
\begin{tabular}{ccc}
\toprule
$\bm{u}$ & $\mathcal{M}(vdn)$ & $\mathcal{M}(its)$\\
\midrule
	$\bm{u}^*$ & $6(1+e_Q)$ & $6(1+e_Q)$\\
    $\bm{\acute{u}}$ & $6$ & $6$ \\ 
    others (inferior actions) & $6(1-\alpha)$& random(-20,6)\\ 
\bottomrule
\end{tabular}
\end{small}
\end{center}
\vskip -0.2in
\end{table}

To evaluate the effectiveness of our method, we compare the ratio of the optimal STN (i.e., the probability of the optimal convergence) of VDN, ITS and GVR as $\epsilon$ grows. The pay-off matrices are generated according to $\mathcal{M}(its)$ in Tab.\ref{rand_mat} over 5 seeds. At each $\epsilon$, 100 times of independent training are executed. As shown in Fig.\ref{exp1.3}(b), the optimal node is not an STN for VDN when $\epsilon>0.8$ since $R_{STN_{opt}}$ becomes 0. While the optimal node is always an STN for ITS, where $R_{STN_{opt}}>0$ always holds. Besides, the optimal STN becomes the unique STN (i.e., $R_{STN_{opt}}=1$) for ITS under large exploration, which consists with the calculation result ($\epsilon_0=0.837$, denoted by the red dash line) of Eq.\ref{e_0}. GVR fails to achieve $R_{STN_{opt}}=1$ under a small $\epsilon$, where the optimal action may be not explored in given training steps.

\begin{figure*}[t]
    \vskip 0.2in
    \begin{center}
    \centerline{\includegraphics[width=\columnwidth*2]{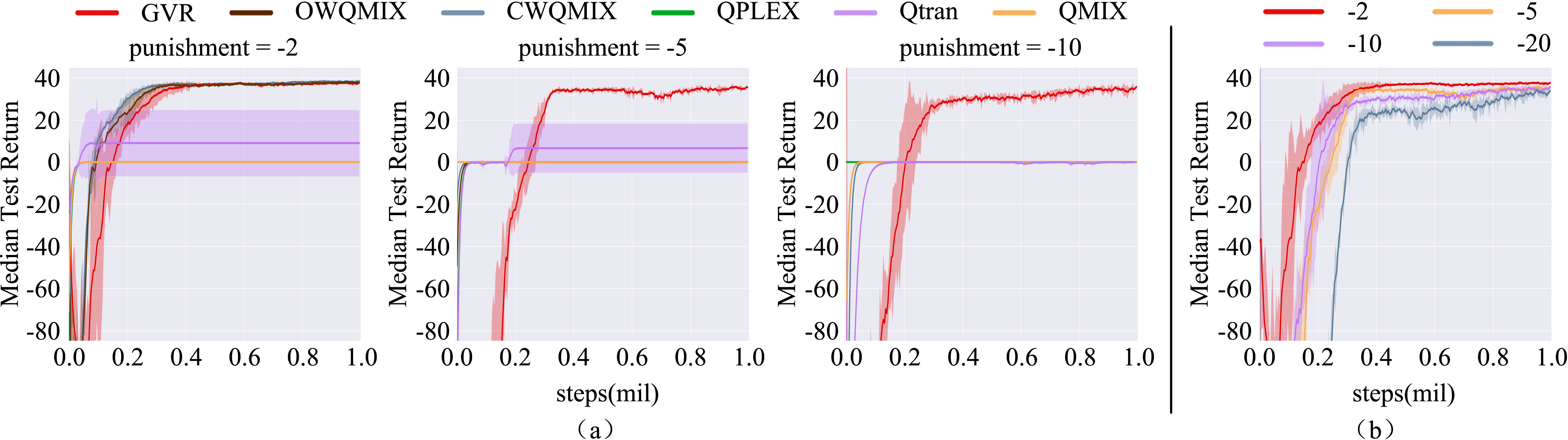}}
    \caption{Experiments on predator-prey. (a) Comparison of GVR and baselines. (b) GVR under different punishments.}\label{pp}
    \end{center}
    \vskip -0.2in
\end{figure*}

\begin{figure}[ht]
    \vskip 0.2in
    \begin{center}
    \centerline{\includegraphics[width=\columnwidth]{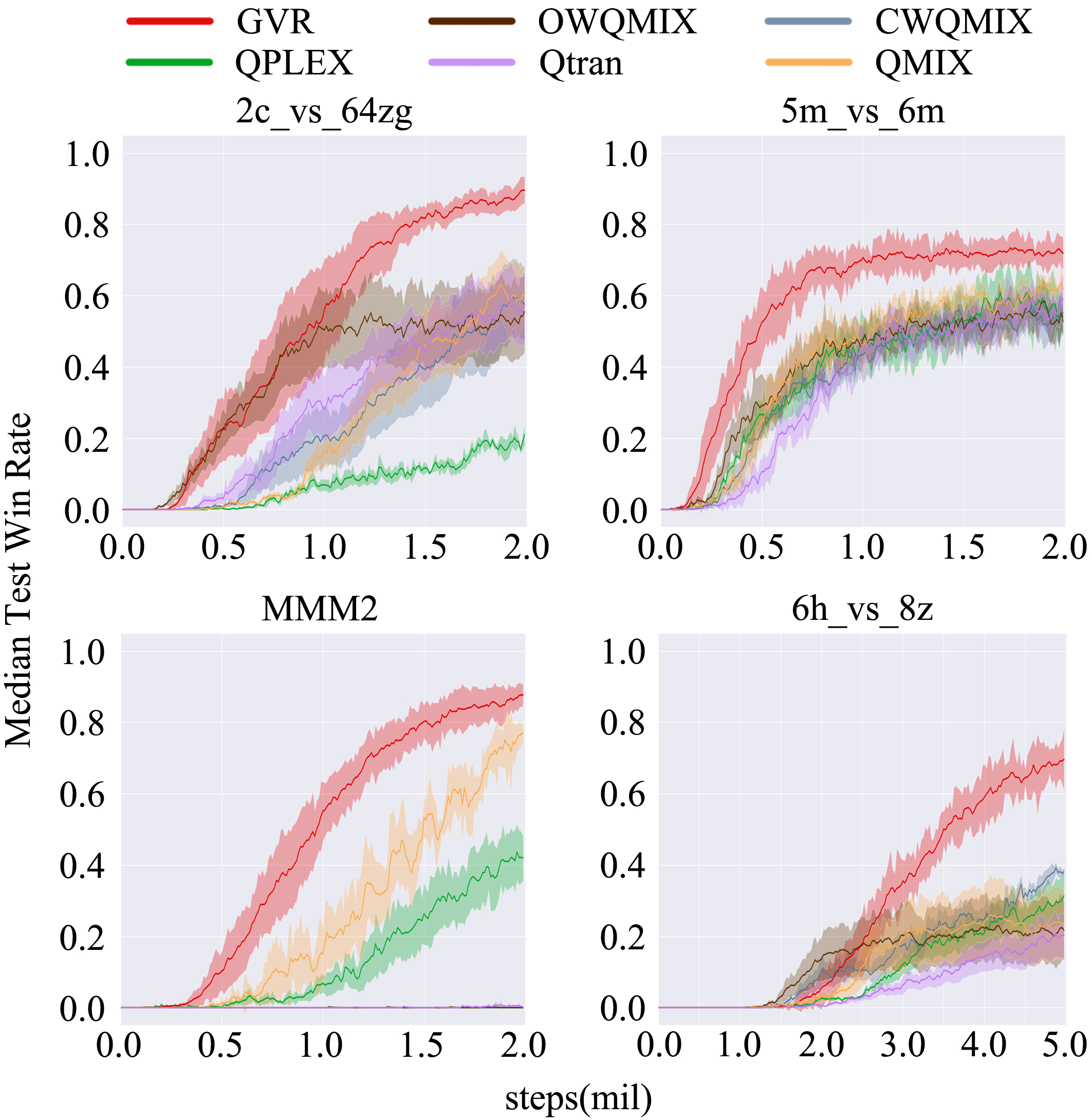}}
    \caption{Comparison of GVR and baselines on SMAC.}
    \label{smac}
    \end{center}
    \vskip -0.2in
\end{figure}

\textbf{GVR vs methods with complete representation capacity.}
We compare GVR to methods with complete representation capacity in matrix games of different scales. Same to $\mathcal{M}(its)$ in Tab.\ref{rand_mat}, the matrix elements for inferior actions are randomly generated over 5 seeds, but the elements of $\bm{u}^*$ and $\bm{\acute{u}}$ are set to $8$ and $6$ respectively. As shown in Fig.\ref{exp1.4}, in the matrix games of size $6^3$ and $12^4$, the representation errors of Qtran, QPLEX and CWQMIX do not decrease in training, which suggests they are unable to learn the complete representation within given steps. We do not measure the representation error of GVR since the target of inferior actions is modified by ITS. GVR is the only method ensuring the optimal consistency  (i.e., median test return $= 8$) on all 3 tasks.

\subsection{Predator-Prey}
Predator-prey is a multi-agent coordinated game, where the predators are trained to capture the preys controlled by random policies. At each time-step, the team of agents receives a positive reward instantly when more than one predators capture a same prey, but is punished with a negative reward when a prey is captured by a single agent. Otherwise, the reward is 0. As a result, the methods suffering from relative overgeneralization (RO) are unable to solve the task, where all agents tend to avoid the preys for fear of the punishment. Our experiments are carried out under 3 punishment values. 

From Fig.\ref{pp}, VDN and QMIX fail in all tasks because both methods suffer from the RO. QPLEX also fails in spite of the complete representation capacity. WQMIX solves the task with the punishment $-2$. However, the heuristic weight $\alpha=0.1$ is insufficient to overcome the RO under large punishments. GVR is able to overcome the RO under all punishments since the true Q values of the inferior actions have little impact on the joint Q values.

\subsection{StarCraft Multi-agent Challenge}
We compare GVR with baselines in challenging tasks of StarCraft multi-agent challenge (SMAC). As shown in Fig.\ref{smac}, GVR shows the best performance. Different from predator-prey with miscoordination punishments, the reward function in SMAC is more reasonable, where the linear and monotonic value decomposition can meet the TGM principle approximately. Therefore, the algorithms with full representation expressiveness capacity (Qtran, QPLEX, WQMIX) do not perform better than QMIX due to the difficulty of complete representation.

We evaluate the effect of the adaptive trade-off, the experiment results are given in Appendix J.2. GVR achieves the trade-off through the ensemble critics, which are independent to the joint Q value function. However, since both the critics and the joint Q value function are evaluations of the policy, the independence brings conflict between two evaluations. As a result, GVR with trade-off performs worse than that without trade-off in some challenging tasks.

\section{Conclusion}
This paper discusses the optimality of credit assignment in value decomposition methods and proposes the optimal consistency. To achieve the optimal consistency, we introduce the TGM principle for linear and monotonic value decomposition. By deriving the expression of the joint Q value function, we draw a transition diagram. According to the diagram, we find the TGM principle can be ensured if the optimal STN is the unique STN. 

By this work, we have a deeper understating about relative overgeneralization (RO). Firstly, we prove theoretically that large exploration do not necessarily solve the RO, it depends on the reward function. Since it is challenging to explore superior actions in tasks with large joint action spaces, efficient exploration is still very important. Secondly, learning a joint Q value function with complete representation capacity is very difficult. Dispensed with complete representation capacity, a biased joint Q value function is sufficient to overcome the RO. However, the heuristic hyper-parameters is unavoidable in previous works since the STNs depend on the task-specific reward function. GVR also learns a biased joint Q value function. But different from previous works, we propose the ITS to remove the dependence of STNs on the true Q values of inferior actions. Besides, GVR achieves an adaptive trade-off between optimality and stability.

\section*{Acknowledgements}

This work was supported in part by National Key R$\&$D Program of China under grant No. 2021ZD0112700, NSFC under grant No.62125305, No.62088102, No.61973246, and the key project of Shaanxi province under grant No.2018ZDCXLGY0605.

\bibliography{example_paper}

\begin{thebibliography}{24}
\providecommand{\natexlab}[1]{#1}
\providecommand{\url}[1]{\texttt{#1}}
\expandafter\ifx\csname urlstyle\endcsname\relax
  \providecommand{\doi}[1]{doi: #1}\else
  \providecommand{\doi}{doi: \begingroup \urlstyle{rm}\Url}\fi

\bibitem[B{\"o}hmer et~al.(2020)B{\"o}hmer, Kurin, and Whiteson]{pp}
B{\"o}hmer, W., Kurin, V., and Whiteson, S.
\newblock Deep coordination graphs.
\newblock In \emph{International Conference on Machine Learning}, pp.\
  980--991. PMLR, 2020.

\bibitem[Foerster et~al.(2018)Foerster, Farquhar, Afouras, Nardelli, and
  Whiteson]{coma}
Foerster, J., Farquhar, G., Afouras, T., Nardelli, N., and Whiteson, S.
\newblock Counterfactual multi-agent policy gradients.
\newblock In \emph{Proceedings of the AAAI Conference on Artificial
  Intelligence}, volume~32, 2018.

\bibitem[Foerster et~al.(2016)Foerster, Assael, De~Freitas, and
  Whiteson]{ctde2}
Foerster, J.~N., Assael, Y.~M., De~Freitas, N., and Whiteson, S.
\newblock Learning to communicate with deep multi-agent reinforcement learning.
\newblock \emph{arXiv preprint arXiv:1605.06676}, 2016.

\bibitem[Guestrin et~al.(2001)Guestrin, Koller, and Parr]{pomdp}
Guestrin, C., Koller, D., and Parr, R.
\newblock Multiagent planning with factored mdps.
\newblock In \emph{NIPS}, volume~1, pp.\  1523--1530, 2001.

\bibitem[Gupta et~al.(2021)Gupta, Mahajan, Peng, B{\"o}hmer, and
  Whiteson]{uneven}
Gupta, T., Mahajan, A., Peng, B., B{\"o}hmer, W., and Whiteson, S.
\newblock Uneven: Universal value exploration for multi-agent reinforcement
  learning.
\newblock In \emph{International Conference on Machine Learning}, pp.\
  3930--3941. PMLR, 2021.

\bibitem[Lowe et~al.(2017)Lowe, Wu, Tamar, Harb, Abbeel, and Mordatch]{maddpg}
Lowe, R., Wu, Y., Tamar, A., Harb, J., Abbeel, P., and Mordatch, I.
\newblock Multi-agent actor-critic for mixed cooperative-competitive
  environments.
\newblock \emph{arXiv preprint arXiv:1706.02275}, 2017.

\bibitem[Mahajan et~al.(2019)Mahajan, Rashid, Samvelyan, and Whiteson]{maven}
Mahajan, A., Rashid, T., Samvelyan, M., and Whiteson, S.
\newblock Maven: Multi-agent variational exploration.
\newblock \emph{arXiv preprint arXiv:1910.07483}, 2019.

\bibitem[Oliehoek \& Amato(2016)Oliehoek and Amato]{decpomdp}
Oliehoek, F.~A. and Amato, C.
\newblock \emph{A concise introduction to decentralized POMDPs}.
\newblock Springer, 2016.

\bibitem[Oliehoek et~al.(2008)Oliehoek, Spaan, and Vlassis]{ctde1}
Oliehoek, F.~A., Spaan, M.~T., and Vlassis, N.
\newblock Optimal and approximate q-value functions for decentralized pomdps.
\newblock \emph{Journal of Artificial Intelligence Research}, 32:\penalty0
  289--353, 2008.

\bibitem[Panait et~al.(2006)Panait, Luke, and Wiegand]{overgene1}
Panait, L., Luke, S., and Wiegand, R.~P.
\newblock Biasing coevolutionary search for optimal multiagent behaviors.
\newblock \emph{IEEE Transactions on Evolutionary Computation}, 10\penalty0
  (6):\penalty0 629--645, 2006.

\bibitem[Rashid et~al.(2018)Rashid, Samvelyan, Schroeder, Farquhar, Foerster,
  and Whiteson]{qmix}
Rashid, T., Samvelyan, M., Schroeder, C., Farquhar, G., Foerster, J., and
  Whiteson, S.
\newblock Qmix: Monotonic value function factorisation for deep multi-agent
  reinforcement learning.
\newblock In \emph{International Conference on Machine Learning}, pp.\
  4295--4304. PMLR, 2018.

\bibitem[Rashid et~al.(2020)Rashid, Farquhar, Peng, and Whiteson]{wqmix}
Rashid, T., Farquhar, G., Peng, B., and Whiteson, S.
\newblock Weighted qmix: Expanding monotonic value function factorisation for
  deep multi-agent reinforcement learning.
\newblock \emph{arXiv preprint arXiv:2006.10800}, 2020.

\bibitem[Samvelyan et~al.(2019)Samvelyan, Rashid, De~Witt, Farquhar, Nardelli,
  Rudner, Hung, Torr, Foerster, and Whiteson]{smac}
Samvelyan, M., Rashid, T., De~Witt, C.~S., Farquhar, G., Nardelli, N., Rudner,
  T.~G., Hung, C.-M., Torr, P.~H., Foerster, J., and Whiteson, S.
\newblock The starcraft multi-agent challenge.
\newblock \emph{arXiv preprint arXiv:1902.04043}, 2019.

\bibitem[Schaul et~al.(2015)Schaul, Quan, Antonoglou, and Silver]{per}
Schaul, T., Quan, J., Antonoglou, I., and Silver, D.
\newblock Prioritized experience replay.
\newblock \emph{arXiv preprint arXiv:1511.05952}, 2015.

\bibitem[Son et~al.(2019)Son, Kim, Kang, Hostallero, and Yi]{qtran}
Son, K., Kim, D., Kang, W.~J., Hostallero, D.~E., and Yi, Y.
\newblock Qtran: Learning to factorize with transformation for cooperative
  multi-agent reinforcement learning.
\newblock In \emph{International Conference on Machine Learning}, pp.\
  5887--5896. PMLR, 2019.

\bibitem[Sunehag et~al.(2017)Sunehag, Lever, Gruslys, Czarnecki, Zambaldi,
  Jaderberg, Lanctot, Sonnerat, Leibo, Tuyls, et~al.]{vdn}
Sunehag, P., Lever, G., Gruslys, A., Czarnecki, W.~M., Zambaldi, V., Jaderberg,
  M., Lanctot, M., Sonnerat, N., Leibo, J.~Z., Tuyls, K., et~al.
\newblock Value-decomposition networks for cooperative multi-agent learning.
\newblock \emph{arXiv preprint arXiv:1706.05296}, 2017.

\bibitem[Vorotnikov et~al.(2018)Vorotnikov, Ermishin, Nazarova, and
  Yuschenko]{robot1}
Vorotnikov, S., Ermishin, K., Nazarova, A., and Yuschenko, A.
\newblock Multi-agent robotic systems in collaborative robotics.
\newblock In \emph{International Conference on Interactive Collaborative
  Robotics}, pp.\  270--279. Springer, 2018.

\bibitem[Wang et~al.(2020)Wang, Ren, Liu, Yu, and Zhang]{qplex}
Wang, J., Ren, Z., Liu, T., Yu, Y., and Zhang, C.
\newblock Qplex: Duplex dueling multi-agent q-learning.
\newblock \emph{arXiv preprint arXiv:2008.01062}, 2020.

\bibitem[Wei et~al.(2018)Wei, Wicke, Freelan, and Luke]{overgene2}
Wei, E., Wicke, D., Freelan, D., and Luke, S.
\newblock Multiagent soft q-learning.
\newblock In \emph{2018 AAAI Spring Symposium Series}, 2018.

\bibitem[Wen et~al.(2020)Wen, Yao, Wang, and Tan]{smix}
Wen, C., Yao, X., Wang, Y., and Tan, X.
\newblock Smix ($\lambda$): Enhancing centralized value functions for
  cooperative multi-agent reinforcement learning.
\newblock In \emph{Proceedings of the AAAI Conference on Artificial
  Intelligence}, volume~34, pp.\  7301--7308, 2020.

\bibitem[Wu et~al.(2020)Wu, Zhou, Liu, Yuan, Wang, Huang, and Wu]{traffic1}
Wu, T., Zhou, P., Liu, K., Yuan, Y., Wang, X., Huang, H., and Wu, D.~O.
\newblock Multi-agent deep reinforcement learning for urban traffic light
  control in vehicular networks.
\newblock \emph{IEEE Transactions on Vehicular Technology}, 69\penalty0
  (8):\penalty0 8243--8256, 2020.

\bibitem[Yang et~al.(2020{\natexlab{a}})Yang, Hao, Chen, Tang, Chen, Hu, Fan,
  and Wei]{qpd}
Yang, Y., Hao, J., Chen, G., Tang, H., Chen, Y., Hu, Y., Fan, C., and Wei, Z.
\newblock Q-value path decomposition for deep multiagent reinforcement
  learning.
\newblock In \emph{International Conference on Machine Learning}, pp.\
  10706--10715. PMLR, 2020{\natexlab{a}}.

\bibitem[Yang et~al.(2020{\natexlab{b}})Yang, Hao, Liao, Shao, Chen, Liu, and
  Tang]{qatten}
Yang, Y., Hao, J., Liao, B., Shao, K., Chen, G., Liu, W., and Tang, H.
\newblock Qatten: A general framework for cooperative multiagent reinforcement
  learning.
\newblock \emph{arXiv preprint arXiv:2002.03939}, 2020{\natexlab{b}}.

\bibitem[Zhang \& Sutton(2017)Zhang and Sutton]{erp}
Zhang, S. and Sutton, R.~S.
\newblock A deeper look at experience replay.
\newblock \emph{arXiv preprint arXiv:1712.01275}, 2017.

\end{thebibliography}
\bibliographystyle{icml2022}

\newpage
\appendix
\onecolumn
\section{Related Work}
\subsection{Relative Overgeneralization and Representation Limitation}
We first explain the relationship between relative overgeneralization and representation limitation. Relative overgeneralization (RO) is firstly proposed in coordination games \citep{overgene1}, where each agent chooses an action without knowing which actions the other agents will choose. All agents receive the same rewards after performing the chosen actions. RO is related to the individual rewards, and it usually occurs in independent learning methods (e.g., independent Q learning) when an agent is punished for the miscoordination of other agents. As a result, the joint policy may converge to a suboptimal equilibrium.

In fully cooperative MARL tasks with credit assignment approaches (e.g., value decomposition), the policy of each agent is evaluated differentially, where RO depends on the credit assignment since it is related to the individual rewards. In linear value decomposition (LVD) or monotonic value decomposition (MVD), a prior linear or monotonic constraint is applied between the joint Q value function and individual utility functions, leading to the representation limitation of the joint Q value function. The representation limitation may result in poor credit assignment, which causes RO.

\subsection{Value Decomposition Methods}
We mainly introduce recent value decomposition methods. Value decomposition is a popular approach for credit assignment in fully cooperative MARL tasks. VDN \citep{vdn} learns a joint Q value function based on a share reward function. In VDN, the joint Q value function is linearly factorized into individual utility functions. By contrast, QMIX \citep{qmix} substitutes the linear factorization with a monotonic factorization, where the weights and bias are produced from the global state through a mixing network. Based on QMIX, SMIX \citep{smix} replaces the TD(0) Q-learning target with a TD($\lambda$) SARSA target. Qatten \citep{qatten} adds an attention network before the mixing network of QMIX. QPD \citep{qpd} decomposes the joint Q value function with the integrated gradient attribution technique, which directly decomposes the joint Q-values along trajectory paths to assign credits for agents. However, due to the representation limitation of the joint Q value function, these methods suffer from the RO.

Some recent works address the RO directly by completing the representation capacity of the joint Q value function. QTRAN \citep{qtran} learns a joint Q value function with complete representation capacity and introduces two soft regularizations to approximate the IGM principle. QPLEX \citep{qplex} achieves the complete representation under IGM principle theoretically through a duelling mixing network, where the complete representation capacity is introduced by the mixing of individual advantage functions. However, as the state space and the joint action space increase exponentially as the number of agents grows, it is impractical to learn the complete representation in complicated MARL tasks, which may result in convergence difficulty and performance deterioration.

The other related works aim to prevent sub-optimal convergences by learning a biased joint Q value function. WQMIX \citep{wqmix} introduces an auxiliary network to distinguish samples with low representation values. By placing a predefined low weight ($\alpha$) on these samples, WQMIX learns a biased joint Q value function which focuses on the representation of actions with good performance. According to our analysis in Appendix F.1, a relatively higher weight on the superior samples helps to eliminate non-optimal STNs under ITS. However, WQMIX can not ensure the optimal consistency for \textit{two reasons}. (1) According to Eq.\ref{mixqfinal} (Appendix C.1), the joint Q value function under MVD depends on the true Q values of all actions in the whole joint action space, which are unavailable and task-specific. As a result, a heuristic weight has to be adopted in WQMIX. (2) According to Tab.\ref{tabxx} (Appendix F.2), the required weight ($\alpha=\frac{1}{w_0}$) in WQMIX is very small under ITS, which introduces instability in training. Although WQMIX eliminates all non-optimal STNs theoretically when $\alpha=0$, the optimal STN is unstable (i.e., the optimal node can not keep self-transition) because the joint Q values of actions with low representation values have not been trained.

MAVEN \citep{maven} focuses on the poor exploration that arises from the representation limitation and introduces a latent space for hierarchical control, which achieves temporally extended exploration. UneVEn \citep{uneven} solves a target task by learning a set of related tasks simultaneously with a linear decomposition of universal successor features, which improves the joint exploration. Both methods improve the joint exploration, which helps to eliminate non-optimal STNs. However, they can not ensure the optimal consistency for \textit{two reasons}. (1) According to Fig.4 (Section 5.1), improving exploration does not necessary ensure the optimal consistency. Instead, it depends on the reward function. (2) According to Eq.9 in Section 4.1, to eliminate non-optimal STNs, the required exploration rate approximates 1 under ITS, which is inapplicable in tasks with long episodes.

Besides, as discussed in Appendix H.2, excessive pursuit for optimality decreases the stability. Without trade-off between optimality and stability, the methods which approximate the optimal consistency (e.g., WQMIX under extreme small weights and QPLEX) suffer from the risk of instability.

\begin{table}[t]
\caption{Comparison between GVR and related works.}
\label{igmtgm}
\vskip 0.15in
\begin{center}
\begin{small}
\begin{sc}
\begin{tabular}{lcccr}
\toprule
Methods & \makecell[c]{Ensures \\IGM principle} & \makecell[c]{Ensures \\ TGM principle} & \makecell[c]{trade-off between\\optimality and stability}\\
\midrule
			IQL & $\times$ & $\times$ & $\times$\\
            VDN & \color{red}$\surd$ & $\times$ & $\times$\\ 
            QMIX & \color{red}$\surd$& $\times$ & $\times$\\ 
            SMIX & \color{red}$\surd$& $\times$ & $\times$\\ 
            Qatten & \color{red}$\surd$ & $\times$ & $\times$\\ 
            QDP  & $\times$ & $\times$ & $\times$\\ 
            QTRAN & $\times$ & \color{red}$\surd$ & $\times$\\ 
            MAVEN  & \color{red}$\surd$ & $\times$ & $\times$\\ 
            UneVEn & \color{red}$\surd$ & $\times$ & $\times$\\ 
            WQMIX  & \color{red}$\surd$ & $\times$ & $\times$\\
            QPLEX & \color{red}$\surd$ & \color{red}$\surd$ & $\times$\\
            GVR(ours) & \color{red}$\surd$ & \color{red}$\surd$& \color{red}$\surd$\\
\bottomrule
\end{tabular}
\end{sc}
\end{small}
\end{center}
\vskip -0.2in
\end{table}

\section{Joint Q Value Function of LVD in Awo-agent Cooperation}
Consider a two-agent fully cooperative task with LVD. The joint Q value function $Q(u^1_i,u^2_j,\bm{\tau})$ is linearly factorized into two utility functions as
\begin{equation}
Q(u^1_i,u^2_j,\bm{\tau})=\mathcal{U}^1(u^1_i,\tau^1)+\mathcal{U}^2(u^2_j,\tau^2)
\end{equation} 
where $u^1_i, u^2_j\in \{u_1,\cdots,u_m\}$ are the actions of agents 1 and 2, respectively. $\{u_1,\cdots,u_m\}$ is the discrete individual action space. Specially, the greedy actions of two agents are denoted by $\acute{u}^1$ and $\acute{u}^2$. For briefness, we denote $\mathcal{U}^a(u^a_i,\tau^a)$ and the true Q value function $\mathcal{Q}(s,u^1_i,u^2_j)$ with $\mathcal{U}^a_i(a\in \{1,2 \})$ and $\mathcal{Q}_{ij}$, respectively. Since the utility functions are trained to approximate different true Q values in different combinations, under $\epsilon-$greedy visitation, we have
\begin{equation}
\begin{aligned}
\mathcal{U}_i^1 = \frac{\epsilon}{m}\sum^m_{k=1}(\mathcal{Q}_{ik}-\mathcal{U}_k^2) + (1-\epsilon)(\mathcal{Q}_{i\acute{j}}-\mathcal{U}^2_{\acute{j}})\\
\mathcal{U}_j^2 = \frac{\epsilon}{m}\sum^m_{k=1}(\mathcal{Q}_{kj}-\mathcal{U}_k^1) + (1-\epsilon)(\mathcal{Q}_{\acute{i}j}-\mathcal{U}^1_{\acute{i}})\label{UaUb}
\end{aligned}
\end{equation} 
The sum of two agents' utility functions over all actions equals to
\begin{equation}
\begin{aligned}
\sum_{i=1}^m{\mathcal{U}_i^1} + \sum_{j=1}^m{\mathcal{U}_j^2}=\frac{\epsilon}{m}\left[\sum^m_{i=1}\sum_{k=1}^m{\mathcal{Q}_{ik}}+\sum^m_{j=1}\sum_{k=1}^m{\mathcal{Q}_{kj}}-m\sum_{k=1}^m(\mathcal{U}_k^1+\mathcal{U}_k^2)\right]+(1-\epsilon)\left[\sum_{i=1}^n(\mathcal{Q}_{i\acute{j}}+\mathcal{Q}_{\acute{i}j})-m(\mathcal{U}^1_{\acute{i}}+\mathcal{U}^2_{\acute{j}})\right]\label{sumab}
\end{aligned}
\end{equation}
Notice that $\mathcal{U}^1_{\acute{i}}+\mathcal{U}^2_{\acute{j}}=Q(\acute{u}^1,\acute{u}^2,\bm{\tau})$, and $\sum^m_{i=1}\sum_{k=1}^m{\mathcal{Q}_{ik}}=\sum^m_{j=1}\sum_{k=1}^m{\mathcal{Q}_{kj}}=\sum_{i=1}^m\sum_{j=1}^m \mathcal{Q}_{ij}$, we have
\begin{equation}
\begin{aligned}
\sum_{k=1}^m(\mathcal{U}_k^1+\mathcal{U}_k^2)=\frac{2\epsilon}{m(1+\epsilon)}\sum_{i=1}^m\sum_{j=1}^m \mathcal{Q}_{ij}+
\frac{1-\epsilon}{1+\epsilon}\sum_{k=1}^m(\mathcal{Q}_{\acute{i} k}+\mathcal{Q}_{k \acute{j}})-
\frac{m(1-\epsilon)}{1+\epsilon}Q(\acute{u}^1,\acute{u}^2,\bm{\tau})\label{sumUaUb}
\end{aligned}
\end{equation}
According to Eq.\ref{UaUb} and Eq.\ref{sumUaUb}, for $\forall i,j\in\left[1,m\right]$, the joint Q value function equals to
\begin{equation}
\begin{aligned}
Q(u^1_i,u^2_j,\bm{\tau})=&\mathcal{U}_i^1+\mathcal{U}_j^2
=\frac{\epsilon}{m}\left[\sum_{k=1}^m(\mathcal{Q}_{ik}+\mathcal{Q}_{kj})-\sum_{k=1}^m(\mathcal{U}_k^1+\mathcal{U}_k^2)\right]+(1-\epsilon)(\mathcal{Q}_{\acute{i} j}+\mathcal{Q}_{i \acute{j}}-Q_{\acute{i}\acute{j}})\\
=&\frac{\epsilon}{m}\sum_{k=1}^m(\mathcal{Q}_{ik}+\mathcal{Q}_{kj})+(1-\epsilon)(\mathcal{Q}_{\acute{i} j}+\mathcal{Q}_{i \acute{j}})-\frac{2\epsilon^2}{m^2(1+\epsilon)}\sum_{i=1}^m\sum_{j=1}^m \mathcal{Q}_{ij}\\
&-\frac{\epsilon(1-\epsilon)}{m(1+\epsilon)}\sum_{k=1}^m(\mathcal{Q}_{\acute{i} k}+\mathcal{Q}_{k\acute{j}})-\frac{1-\epsilon}{1+\epsilon}Q(\acute{u}^1,\acute{u}^2,\bm{\tau}) \label{qij}
\end{aligned}
\end{equation}
Notice that $Q(u^1_i,u^2_j,\bm{\tau})$ is related to the joint greedy Q value $Q(\acute{u}^1,\acute{u}^2,\bm{\tau})$. To remove it, substituting $u^1_i=\acute{u}^1$ and $u^2_j=\acute{u}^2$ into Eq.\ref{qij}, we have
\begin{equation}
\begin{aligned}
Q(\acute{u}^1,\acute{u}^2,\bm{\tau})=\frac{\epsilon^2}{m}\sum_{k=1}^m(\mathcal{Q}_{\acute{i}k}+\mathcal{Q}_{k\acute{j}})-\frac{\epsilon^2}{m^2}\sum_{i=1}^m\sum_{j=1}^m \mathcal{Q}_{ij}+(1-\epsilon^2)\mathcal{Q}_{\acute{i}\acute{j}}\label{q*}
\end{aligned}
\end{equation}
Substituting Eq.\ref{q*} into Eq.\ref{qij}, $Q(u^1_i,u^2_j,\bm{\tau})$ can be represented by true Q values as
\begin{equation}
\begin{aligned}
Q(u^1_i,u^2_j,\bm{\tau})=&\frac{\epsilon}{m}\sum_{k=1}^m(\mathcal{Q}_{ik}+\mathcal{Q}_{kj})+(1-\epsilon)(\mathcal{Q}_{\acute{i} j}+\mathcal{Q}_{i\acute{j}})-\frac{\epsilon^2}{m^2}\sum_{i=1}^m\sum_{j=1}^m \mathcal{Q}_{ij}\\
&-\frac{\epsilon(1-\epsilon)}{m}\sum_{k=1}^m(\mathcal{Q}_{\acute{i} k}+\mathcal{Q}_{k\acute{j}})-(1-\epsilon)^2\mathcal{Q}_{\acute{i}\acute{j}}\label{qfinalapp}
\end{aligned}
\end{equation}

\section{Joint Q Value Function of MVD in Awo-agent Cooperation}
\subsection{Derivation}
For two-agent monotonic value decomposition, the joint Q value function is decomposed as $Q_{ij}=\omega_1(s)\mathcal{U}_i^1+\omega_2(s)\mathcal{U}_j^2+V(s)$, where $V(s)$ is a bias. $\omega_1(s)$ and $\omega_2(s)$ are the coefficients of $\mathcal{U}_i^1$ and $\mathcal{U}_j^2$, respectively. For brevity, we omit their inputs. Referring to Eq.\ref{UaUb}, the individual utility functions with coefficients equal to
\begin{equation}
\begin{aligned}
\omega_1\mathcal{U}_i^1 &= \frac{\epsilon}{m}\sum^m_{k=1}\left[\mathcal{Q}_{ik}-\omega_2\mathcal{U}_k^2-V\right] + (1-\epsilon)\left[\mathcal{Q}_{i \acute{j}}-\omega_2\mathcal{U}^2_{\acute{j}}-V\right]\\
&=\frac{\epsilon}{m}\sum^m_{k=1}\left[\mathcal{Q}_{ik}-\omega_2\mathcal{U}_k^2\right] + (1-\epsilon)\left[\mathcal{Q}_{i \acute{j}}-\omega_2\mathcal{U}^2_{\acute{j}}\right]-V\\
\omega_2\mathcal{U}_j^2 &= \frac{\epsilon}{m}\sum^m_{k=1}\left[\mathcal{Q}_{kj}-\omega_1\mathcal{U}_k^1-V\right] + (1-\epsilon)\left[\mathcal{Q}_{\acute{i} j}-\omega_1\mathcal{U}^1_{\acute{i}}-V\right]\\
&=\frac{\epsilon}{m}\sum^m_{k=1}\left[\mathcal{Q}_{ik}-\omega_2\mathcal{U}_k^2\right] + (1-\epsilon)\left[\mathcal{Q}_{i\acute{j}}-\omega_2\mathcal{U}^2_{\acute{j}}\right]-V\label{mixUaUb}
\end{aligned}
\end{equation} 
Notice $\omega_1(s)$, $\omega_2(s)$ and $V(s)$ are independent of the actions. Referring to the derivation of Eq.\ref{sumUaUb}, we have
\begin{equation}
\begin{aligned}
\sum_{k=1}^m(\omega_1\mathcal{U}_k^1+\omega_2\mathcal{U}_k^2)=\frac{2\epsilon}{m(1+\epsilon)}\sum_{i=1}^m\sum_{j=1}^m \mathcal{Q}_{ij}+
\frac{1-\epsilon}{1+\epsilon}\sum_{k=1}^m(\mathcal{Q}_{\acute{i} k}+\mathcal{Q}_{k\acute{j}})-
\frac{m(1-\epsilon)}{1+\epsilon}Q(\acute{u}^1,\acute{u}^2,\bm{\tau})-mV\label{mixsumUaUb}
\end{aligned}
\end{equation}
According to Eq.\ref{mixUaUb} and Eq.\ref{mixsumUaUb}, we have
\begin{equation}
\begin{aligned}
Q(u^1_i,u^2_j,\bm{\tau})=&\frac{\epsilon}{m}\sum_{k=1}^m(\mathcal{Q}_{ik}+\mathcal{Q}_{kj})+(1-\epsilon)(\mathcal{Q}_{\acute{i}j}+\mathcal{Q}_{i\acute{j}})-\frac{2\epsilon^2}{m^2(1+\epsilon)}\sum_{i=1}^m\sum_{j=1}^m \mathcal{Q}_{ij}\\
&-\frac{1-\epsilon}{1+\epsilon}Q(\acute{u}^1,\acute{u}^2,\bm{\tau})-\frac{\epsilon(1-\epsilon)}{m(1+\epsilon)}\sum_{k=1}^m(\mathcal{Q}_{\acute{i} k}+\mathcal{Q}_{k\acute{j}})\label{mixqij}
\end{aligned}
\end{equation}
In order to remove $Q(\acute{u}^1,\acute{u}^2,\bm{\tau})$ from Eq.\ref{mixqij}, let $u^1_i=\acute{u}^1$ and $u^2_j=\acute{u}^2$.
\begin{equation}
\begin{aligned}
Q(\acute{u}^1,\acute{u}^2,\bm{\tau}) =\frac{\epsilon^2}{m}\sum_{k=1}^m(\mathcal{Q}_{\acute{i} k}+\mathcal{Q}_{k\acute{j}})-\frac{\epsilon^2}{m^2}\sum_{i=1}^m\sum_{j=1}^m \mathcal{Q}_{ij}+(1-\epsilon^2)\mathcal{Q}_{\acute{i}\acute{j}}\label{mixq*}
\end{aligned}
\end{equation}
Substituting Eq.\ref{mixq*} into Eq.\ref{mixqij}, we have
\begin{equation}
\begin{aligned}
Q(u^1_i,u^2_j,\bm{\tau})=&\frac{\epsilon}{m}\sum_{k=1}^m(\mathcal{Q}_{ik}+\mathcal{Q}_{kj})+(1-\epsilon)(\mathcal{Q}_{\acute{i} j}+\mathcal{Q}_{i\acute{j}})-\frac{\epsilon^2}{m^2}\sum_{i=1}^m\sum_{j=1}^m \mathcal{Q}_{ij}\\
&-\frac{\epsilon(1-\epsilon)}{m}\sum_{k=1}^m(\mathcal{Q}_{\acute{i} k}+\mathcal{Q}_{k\acute{j}})-(1-\epsilon)^2\mathcal{Q}_{\acute{i}\acute{j}}\label{mixqfinal}
\end{aligned}
\end{equation}
Eq.\ref{mixqfinal} is the same as Eq.\ref{qfinalapp}, which indicates LVD and MVD share \textit{the same expression} of the joint Q value function.

\subsection{Verification}
\begin{figure}[ht]
    \vskip 0.2in
    \begin{center}
    \centerline{\includegraphics[width=\columnwidth*4/13]{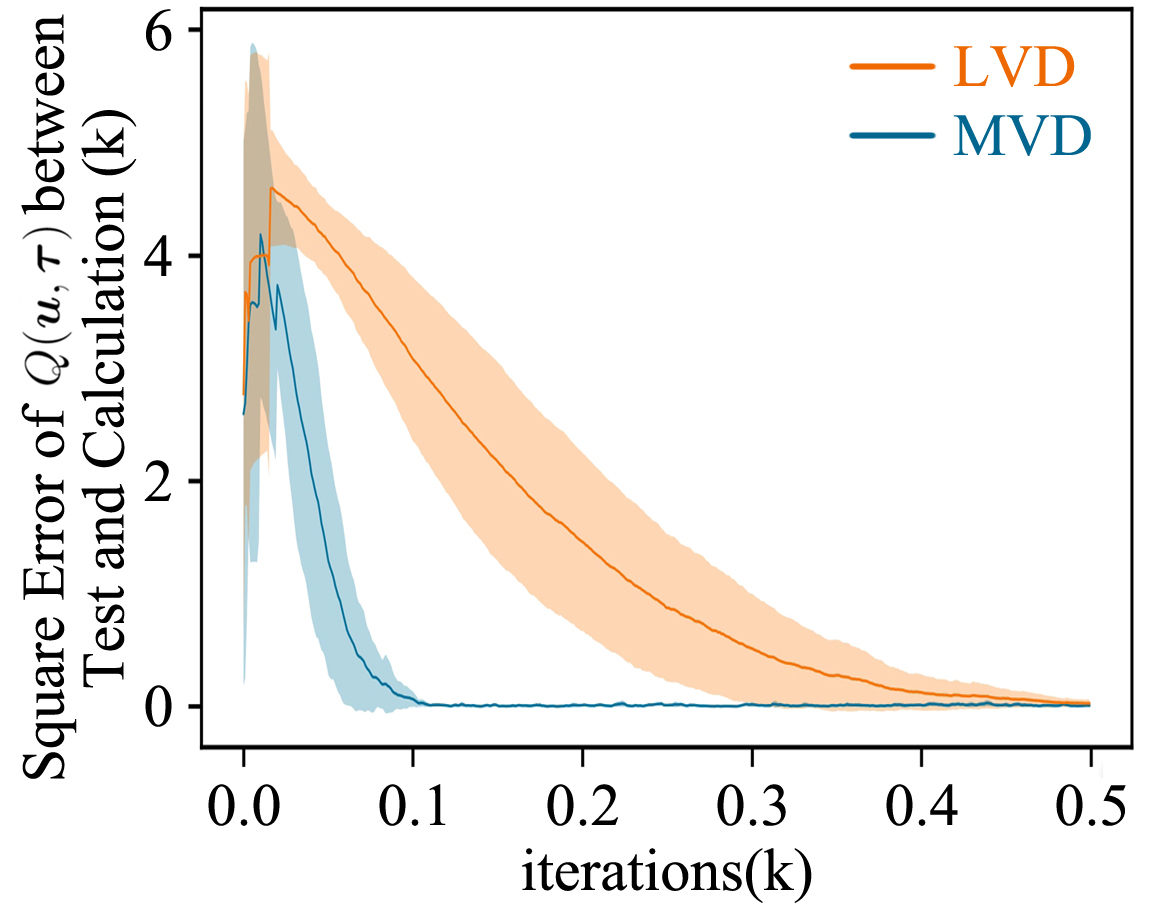}}
    \caption{Square error of $Q(\bm{u},\bm{\tau})$ between test and calculation.}\label{exp1.0}
    \end{center}
    \vskip -0.2in
\end{figure}

We verify the derived joint Q value function (i.e., Eq.\ref{qfinalapp}) for LVD and MVD in a two-agent matrix game, where the payoff matrix is shown in Tab.\ref{tab2}(a). Since the episode length is $1$, an mlp shared by two agents is adopted as the policy network. The policy network is trained for 500 iterations (100 episodes per iteration) over 5 seeds under $\epsilon=0.2$. 
As shown in Tab.\ref{tab2}(b) and Tab.\ref{tab2}(c), there are two STNs. The test results of joint Q values for LVD and MVD indicates LVD and MVD share the same expression of the joint Q value function. Besides, we measure the square error of joint Q values between test ($Q_{test}(\bm{u},\bm{\tau})$) and calculation ($Q_{cal}(\bm{u},\bm{\tau})$), i.e., $\sum_{\bm{u}}^{\bm{U}}\left[Q_{cal}(\bm{u},\bm{\tau})-Q_{test}(\bm{u},\bm{\tau})\right]^2$. The result is shown in Fig.\ref{exp1.0}, where MVD converges faster than LVD.

\begin{minipage}{1.0\linewidth}
\tabcaption{Verification of the joint Q value function (Eq.\ref{qfinalapp}) for two-agent LVD and MVD. (a) The pay-off matrix. (b),(c) Comparison between test and calculation joint Q values for LVD and MVD on two STNs. We mark the calculation result, test result of LVD and MVD with $(C)$, $(L)$ and $(M)$ respectively. The greedy policies are marked with pink backgrounds.}\label{tab2}
\vskip 0.15in
\begin{small}
\begin{center}
	\begin{tabular}{|>{\centering\arraybackslash}p{0.92cm}|>{\centering\arraybackslash}p{0.92cm}|>{\centering\arraybackslash}p{0.92cm}|}
        		\hline
            8 & -12 & -12\\ \hline
            -12 & 0& 0\\ \hline
            -12 & 0& 6\\ \hline
        \end{tabular}
        \vspace{0.1cm}
        \\(a)
\end{center}
\begin{multicols}{2}
\begin{center}
		\begin{tabular}{|c|c|c|}
					\hline
				\cellcolor{pink} \makecell[c]{7.40$(C)$\\7.38$\pm$0.02$(L)$\\
				7.39$\pm$0.07$(M)$} & \makecell[c]{-8.33$(C)$\\-8.27$\pm$0.12$(L)$\\
				-8.56$\pm$0.06$(M)$} & \makecell[c]{-7.93$(C)$\\-7.86$\pm$0.13$(L)$\\
				-7.88$\pm$0.08$(M)$}\\ \hline
				\makecell[c]{-8.33$(C)$\\-8.33$\pm$0.18$(L)$\\-8.18$\pm$0.12$(M)$} & \makecell[c]{-24.06$(C)$\\-23.87$\pm$0.09$(L)$\\-24.51$\pm$0.06$(M)$} & \makecell[c]{-23.66$(C)$\\-23.56$\pm$0.17$(L)$\\-23.41$\pm$0.17$(M)$}\\ \hline
				\makecell[c]{-7.93$(C)$\\-7.93$\pm$0.09$(L)$\\-8.08$\pm$0.14$(M)$}& 
				\makecell[c]{-23.66$(C)$\\-23.53$\pm$0.12$(L)$\\-24.44$\pm$0.17$(M)$} & \makecell[c]{-23.26$(C)$\\-23.19$\pm$0.20$(L)$\\-23.36$\pm$0.23$(M)$}\\ \hline
			\end{tabular}
			\vspace{0.1cm}
			\\(b)
\end{center}
\begin{center}
\begin{tabular}{|c|c|c|}
	\hline
	\makecell[c]{-24.38$(C)$\\-24.34$\pm$0.25$(L)$\\-23.98$\pm$0.18$(M)$} & \makecell[c]{-14.52$(C)$\\-14.52$\pm$0.11$(L)$\\-14.34$\pm$0.08$(M)$} & \makecell[c]{-9.32$(C)$\\-9.43$\pm$0.15$(L)$\\-9.20$\pm$0.05$(M)$}\\ \hline
	\makecell[c]{-14.52$(C)$\\-14.47$\pm$0.18$(L)$\\-14.23$\pm$0.09$(M)$} & \makecell[c]{-4.65$(C)$\\-4.65$\pm$0.11$(L)$\\-4.59$\pm$0.06$(M)$} & \makecell[c]{0.55$(C)$\\0.54$\pm$0.08$(L)$\\0.55$\pm$0.05$(M)$}\\ \hline
	\makecell[c]{-9.32$(C)$\\-9.28$\pm$0.21$(L)$\\-9.07$\pm$0.14$(M)$} & \makecell[c]{0.55$(C)$\\0.56$\pm$0.12$(L)$\\0.57$\pm$0.06$(M)$} & \cellcolor{pink} \makecell[c]{5.75$(C)$\\5.75$\pm$0.09$(L)$\\5.72$\pm$0.05$(M)$}\\ \hline
	\end{tabular}
	\vspace{0.1cm}
	\\(c)
\end{center}
\vskip -0.2in
\end{multicols}
\end{small}
\end{minipage}

\section{Derivation of $Q(\bm{u}_s,\bm{\tau})-Q(\bm{\acute{u}},\bm{\tau})$ under ITS}
\subsection{Examples of the ITS Target}
\begin{figure}[h]
	\centering
	\includegraphics[scale=0.18]{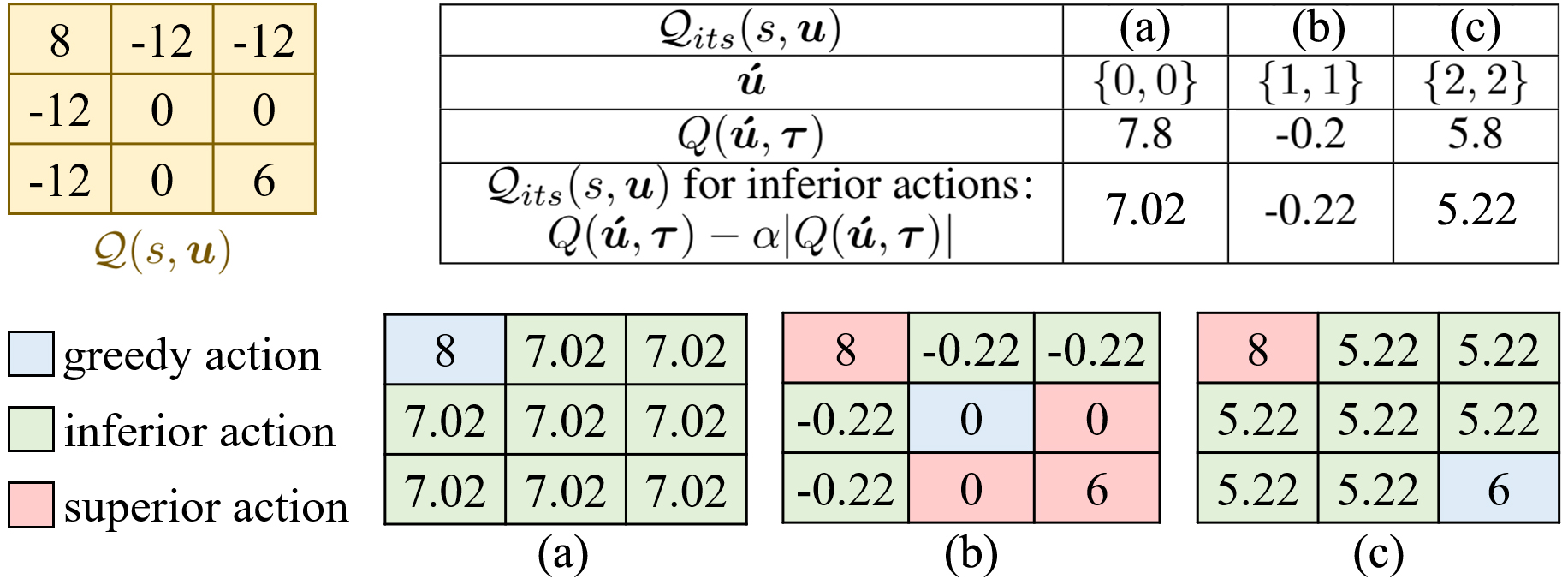}
	\caption{Examples of the ITS target under different $\bm{\acute{u}}$, where $e_{Q0}=0.1$ and $\alpha=0.1$. The true Q values are given in the yellow table.}
	\label{itsexam}
\end{figure}

Examples of the ITS target under different greedy actions are shown in Fig.\ref{itsexam}.

\subsection{Derivation 1}
Given the greedy action $\{\acute{u}^1,\cdots,\acute{u}^n\}=\bm{\acute{u}}$ and any other action $\{u^1_s,\cdots,u^n_s\}=\bm{u}_s\neq \bm{\acute{u}}$, here we regard all actions except $\bm{u}_s$ and $\bm{\acute{u}}$ as inferior actions. Discussions under multiple superior actions is provided in Appendix H.1. We first consider the hardest exploration case, where $\forall a\in[1,n]$, $u^a_s\neq \acute{u}^{a}$. The derivation under non-hardest exploration cases is provided in Appendix D.4. For simplicity, assuming $Q(\bm{\acute{u}},\bm{\tau})>0$, where $\mathcal{Q}_{its}(s, \bm{u})=(1-\alpha)Q(\bm{\acute{u}},\bm{\tau})$ for inferior actions, the utility function of individual action $u^a_s(a\in[1,n])$ is consist of two parts
\begin{equation}
\begin{aligned}
\mathcal{U}^a(u^a_s,\tau^a) =& (1-\eta_1)\left[ (1-\alpha) Q(\bm{\acute{u}},\bm{\tau})-\sum_{k}^{m^{n-1}-1}\left[ \frac{p(u^a_s,u^{-a}_k)}{p(u^a_s)-p(\bm{u}_s)} \sum_{i}^{-a}\mathcal{U}^i(u^i_k,\tau^i)\right] \right]\\
&+\eta_1\left[\mathcal{Q}_{its}(s, \bm{u}_s)-\sum_i^{-a} \mathcal{U}^i(u^i_s,\tau^i) \right]
\label{us1}
\end{aligned}
\end{equation}
where $\eta_1=(\frac{\epsilon}{m})^{n-1}$, and $-a$ represents the collection of all agents except agent $a$. $\eta_1$ and $1-\eta_1$ are the proportions of $\bm{u}_s$ and inferior actions in $\sum_{k}^{m^{n-1}}\{u^a_s,u_k^{-a}\}$ respectively, where
$\sum_{k}^{m^{n-1}}\{u^a_s,u_k^{-a}\}$ denotes all joint actions containing $u^a_s$, and $p(u^a_s,u^{-a}_k)$ is the corresponding probability of each joint action. The superscript of the first $\sum$ in Eq.\ref{us1} is $m^{n-1}-1$ because $\bm{u}_s$ is excluded. $p(u^a_s)-p(\bm{u}_s)=\sum_{k}^{m^{n-1}}p(u^a_s,u^{-a}_k)$ is the normalization coefficient, where $\sum_{k}^{m^{n-1}}p(u^a_s,u^{-a}_k)=\frac{\epsilon}{m}-(\frac{\epsilon}{m})^n$. Notice
\begin{equation}
\begin{aligned}
(1-\eta_1)\sum_{k}^{m^{n-1}-1} \left[\frac{p(u^a_s,u^{-a}_k)}{p(u^a_s)-p(\bm{u}_s)} \sum_{i}^{-a}\mathcal{U}^i(u^i_k,\tau^i)\right] + \eta_1\sum_{i}^{-a}\mathcal{U}^i(u^i_k,\tau^i) = \sum_{k}^{m^{n-1}} \left[\frac{p(u^a_s,u^{-a}_k)}{p(u^a_s)} \sum_{i}^{-a}\mathcal{U}^i(u^i_s,\tau^i)\right]
\label{ee1}
\end{aligned}
\end{equation}
Therefore,
\begin{equation}
\begin{aligned}
\mathcal{U}^a(u^a_s,\tau^a) =& (1-\eta_1) (1-\alpha) Q(\bm{\acute{u}},\bm{\tau})-\sum_{k}^{m^{n-1}}\left[ \frac{p(u^a_s,u^{-a}_k)}{p(u^a_s)} \sum_{i}^{-a}\mathcal{U}^i(u^i_k,\tau^i)\right] +\eta_1\mathcal{Q}_{its}(s, \bm{u}_s)\\
\label{us_final1}
\end{aligned}
\end{equation}
The joint Q value function $Q(\bm{u}_s,\bm{\tau})$ can be acquired
\begin{equation}
\begin{aligned}
Q(\bm{u}_s,\bm{\tau}) = \sum_{a=1}^n \mathcal{U}^a(u^a_s,\tau^a) = n(1-\eta_1) (1-\alpha) Q(\bm{\acute{u}},\bm{\tau})- \sum_{a=1}^n \sum_{k}^{m^{n-1}} \left[\frac{p(u^a_s,u^{-a}_k)}{p(u^a_s)} \sum_{i}^{-a}\mathcal{U}^i(u^i_k,\tau^i)\right] +n\eta_1\mathcal{Q}_{its}(s, \bm{u}_s)
\label{Qs1}
\end{aligned}
\end{equation}
Similarly, for the greedy action $\bm{\acute{u}}$, we have 
\begin{equation}
\begin{aligned}
\mathcal{U}^a(\acute{u}^a,\tau^a) =& (1-\eta_2) (1-\alpha) Q(\bm{\acute{u}},\bm{\tau})-\sum_{k}^{m^{n-1}}\left[ \frac{p(\acute{u}^a,u^{-a}_k)}{p(\acute{u}^a)} \sum_{i}^{-a}\mathcal{U}^i(u^i_k,\tau^i)\right] +\eta_2\mathcal{Q}(s, \bm{\acute{u}})\\
\label{u*}
\end{aligned}
\end{equation}
where $\eta_2=(1-\epsilon+\frac{\epsilon}{m})^{n-1}$. As a result,
\begin{equation}
\begin{aligned}
Q(\bm{\acute{u}},\bm{\tau})=\sum_{a=1}^n \mathcal{U}^a(\acute{u}^a,\tau^a)= n(1-\eta_2) (1-\alpha) Q(\bm{\acute{u}},\bm{\tau})- \sum_{a=1}^n \sum_{k}^{m^{n-1}} \left[\frac{p(\acute{u}^a,u^{-a}_k)}{p(\acute{u}^a)} \sum_{i}^{-a}\mathcal{U}^i(u^i_k,\tau^i)\right] +n\eta_2\mathcal{Q}(s, \bm{\acute{u}})\label{Q*1}
\end{aligned}
\end{equation}
Notice $\frac{p(u^{a},u^{-a})}{p(u^{a})}$ is independent of action $u^a$ for decentralized execution, therefore $\frac{p(\acute{u}^a,u^{-a}_k)}{p(\acute{u}^a)}=\frac{p(u^a_s,u^{-a}_k)}{p(u^a_s)}$. Let $\mathcal{Q}(s,\bm{u}_s)=(1+e_Q)\mathcal{Q}(s,\bm{\acute{u}})$ ,according to Eq.\ref{Qs1} and Eq.\ref{Q*1}, we have
\begin{equation}
\begin{aligned}
Q(\bm{u}_s,\bm{\tau})-Q(\bm{\acute{u}},\bm{\tau})= n(\eta_1-\eta_2)\left[\mathcal{Q}(s, \bm{\acute{u}})-(1-\alpha)Q(\bm{\acute{u}},\bm{\tau})\right] + n\eta_1 e_Q\mathcal{Q}(s,\bm{\acute{u}})\label{con_nagentsapp1}
\end{aligned}
\end{equation}
For monotonic value decomposition, Eq.\ref{con_nagentsapp1} also holds since $Q(\bm{\acute{u}}, \bm{\tau})$ and $Q(\bm{u}_s, \bm{\tau})$ do not change. Verification of Eq.\ref{con_nagentsapp1} is provided in the experimental part.

\subsection{Derivation 2}
Given the greedy action $\{\acute{u}^1,\cdots,\acute{u}^n\}=\bm{\acute{u}}$ and any other action $\{u^1_s,\cdots,u^n_s\}=\bm{u}_s\neq \bm{\acute{u}}$, here we regard all actions except $\bm{u}_s$ and $\bm{\acute{u}}$ as inferior actions. Discussions under multiple superior actions is provided in Appendix H.1. We first consider the hardest exploration case, where $\forall a\in[1,n]$, $u^a_s\neq \acute{u}^{a}$. The derivation under non-hardest exploration cases is provided in Appendix D.4. For simplicity, assuming $Q(\bm{\acute{u}},\bm{\tau})>0$, where $\mathcal{Q}_{its}(s, \bm{u})=(1-\alpha)Q(\bm{\acute{u}},\bm{\tau})$ for inferior actions, the utility function of $u^a_s$ equals to
\begin{equation}
\begin{aligned}
\mathcal{U}_{u^a_s}^a=&\left(\frac{\epsilon}{m}\right)^{n-1}\left[C^0_{n-1}(m^{n-1}-1)(1-\alpha)Q(\bm{\acute{u}},\bm{\tau})+ \mathcal{Q}(s, \bm{u}_s)+f_1(\sum^{-a}_o\sum_{i=1}^m\mathcal{U}^o_{u^o_i},\sum_{a=1}^n\mathcal{U}^a_{\acute{u}^a})\right]\\
&+\cdots+\left(\frac{\epsilon}{m}\right)^{n-t}(1-\epsilon)^{t-1}\left[C^{t-1}_{n-1}m^{n-t}(1-\alpha)Q(\bm{\acute{u}},\bm{\tau})+f_t(\sum^{-a}_o\sum_{i=1}^m\mathcal{U}^o_{u^o_i},\sum_{a=1}^n\mathcal{U}^a_{\acute{u}^a})\right]+\cdots\\
&+(1-\epsilon)^{n-1}\left[ C^{n-1}_{n-1}(1-\alpha)Q(\bm{\acute{u}},\bm{\tau})+f_n(\sum^{-a}_o\sum_{i=1}^m\mathcal{U}^o_{u^o_i},\sum_{a=1}^n\mathcal{U}^a_{\acute{u}^a})\right]\\
=&(1-\alpha)Q(\bm{\acute{u}},\bm{\tau})+\left(\frac{\epsilon}{m}\right)^{n-1} \left[\mathcal{Q}(s,\bm{u}_s)-(1-\alpha)Q(\bm{\acute{u}},\bm{\tau})\right]+f_{total}(\sum^{-a}_o\sum_{i=1}^m\mathcal{U}^o_{u^o_i},\sum_{a=1}^n\mathcal{U}^a_{\acute{u}^a})\label{un}
\end{aligned}
\end{equation}
where $-a$ represents the collection of all agents except agent $a$. $f_t(t\in[1,n])$ and $f_{total}$ are mappings from $\{\sum^{-a}_o\sum_{i=1}^m\mathcal{U}^o_{u^o_i},\sum_{a=1}^n\mathcal{U}^a_{\acute{u}^a}\}$ to $\mathbb{R}$. The joint $Q$ value of $\bm{u}_s$ equals to
\begin{equation}
\begin{aligned}
Q(\bm{u}_s, \bm{\tau})=\sum_{a=1}^n \mathcal{U}_{u^a_s}^a= n(1-\alpha)Q(\bm{\acute{u}},\bm{\tau})+n\left(\frac{\epsilon}{m}\right)^{n-1}\left[\mathcal{Q}(s, \bm{u}_s)-(1-\alpha)Q(\bm{\acute{u}},\bm{\tau})\right]+nf_{total}(\sum^{-a}_o\sum_{i=1}^m\mathcal{U}^o_{u^o_i},\sum_{a=1}^n\mathcal{U}^a_{\acute{u}^a})\label{Qn}
\end{aligned}
\end{equation}
Next we calculate the joint Q value of $\bm{\acute{u}}$. The utility function of the individual greedy action $\acute{u}^a(a\in[1,n])$ equals to
\begin{equation}
\begin{aligned}
\mathcal{U}_{\acute{u}^a}^a=(1-\alpha)Q(\bm{\acute{u}},\bm{\tau})+
(1-\epsilon+\frac{\epsilon}{m})^{n-1} \left[\mathcal{Q}(s,\bm{\acute{u}})-(1-\alpha)Q(\bm{\acute{u}},\bm{\tau})\right] +f_{total}(\sum^{-a}_o\sum_{i=1}^m\mathcal{U}^o_{u^o_i},\sum_{a=1}^n\mathcal{U}^a_{\acute{u}^a})
\end{aligned}
\end{equation}
The joint $Q$ value of $\bm{\acute{u}}$ equals to
\begin{equation}
\begin{aligned}
Q(\bm{\acute{u}},\bm{\tau})=&\sum_{a=1}^n \mathcal{U}_{\acute{u}^a}^a\\
=&n(1-\alpha)Q(\bm{\acute{u}},\bm{\tau})+n(1-\epsilon+\frac{\epsilon}{m})^{n-1} \left[\mathcal{Q}(s,\bm{\acute{u}})-(1-\alpha)Q(\bm{\acute{u}},\bm{\tau})\right] +nf_{total}(\sum^{-a}_o\sum_{i=1}^m\mathcal{U}^o_{u^o_i},\sum_{a=1}^n\mathcal{U}^a_{\acute{u}^a})\label{Qn*}
\end{aligned}
\end{equation}
Let $\eta_1=(\frac{\epsilon}{m})^{n-1}$, $\eta_2=(1-\epsilon+\frac{\epsilon}{m})^{n-1}$, and $\mathcal{Q}(s,\bm{u}_s)=(1+e_Q)\mathcal{Q}(s,\bm{\acute{u}})$, we have
\begin{equation}
\begin{aligned}
Q(\bm{u}_s, \bm{\tau})-Q(\bm{\acute{u}},\bm{\tau})&= n\eta_1\left[(1+e_Q)\mathcal{Q}(s, \bm{\acute{u}})-(1-\alpha)Q(\bm{\acute{u}},\bm{\tau})\right]-n\eta_2\left[\mathcal{Q}(s, \bm{\acute{u}})-(1-\alpha)Q(\bm{\acute{u}},\bm{\tau})\right]\\
&= n(\eta_1-\eta_2)\left[\mathcal{Q}(s, \bm{\acute{u}})-(1-\alpha)Q(\bm{\acute{u}},\bm{\tau})\right] + n\eta_1 e_Q\mathcal{Q}(s,\bm{\acute{u}})
\label{con_nagentsapp}
\end{aligned}
\end{equation}
which \textit{consists with} Eq.\ref{con_nagentsapp1} in Derivation 1.

\subsection{Non-hardest Exploration Cases}
In Proof 1 and Proof 2, we only consider the hardest exploration case, where $\forall a\in[1,n]$, $ u^a_s \neq \acute{u}^a$. In this subsection, we derive $Q(\bm{u}_s, \bm{\tau})-Q(\bm{\acute{u}},\bm{\tau})$ in general cases where 
\begin{equation}
u^a_s \left\{
\begin{array}{lr}
 = \acute{u}^a \ \ \ \ \ \ \ \ \ \ \ \ \ \ \ \ a\in \mathcal{L} \\ 
 \neq \acute{u}^a \ \ \ \ \ \ \ \ \ \ \ \ \ \ \ \ others
\end{array}
\right.	\label{u_gen}
\end{equation}
$\mathcal{L}\in\{1,\cdots,n\}$ and $\mathcal{L}\neq\emptyset$. Assuming $\mathcal{Q}(s, \bm{\acute{u}})>0$, i.e., $\mathcal{Q}_{its}(s, \bm{u})=(1-\alpha)Q(\bm{\acute{u}},\bm{\tau})$ for inferior samples. When $a\in\mathcal{L}$, the utility function of individual action $u^a_s$ is consist of three parts
\begin{equation}
\begin{aligned}
\mathcal{U}^a(u^a_s,\tau^a)=&\eta_1'\left[\mathcal{Q}_{its}(s, \bm{u}_s)-\sum_i^{-a} \mathcal{U}^i(u^i_s,\tau^i) \right]+\eta_2\left[\mathcal{Q}(s, \bm{\acute{u}})-\sum_i^{-a} \mathcal{U}^i(\acute{u}^i,\tau^i) \right]\\
&+ (1-\eta_1'-\eta_2)\left[ (1-\alpha) Q(\bm{\acute{u}},\bm{\tau})-\sum_{k}^{m^{n-1}-2}\left[ \frac{p(u^a_s,u^{-a}_k)}{p(u^a_s)-p(\bm{u}_s)} \sum_{i}^{-a}\mathcal{U}^i(u^i_k,\tau^i)\right] \right] \label{ugeneral}
\end{aligned}
\end{equation}
where $\eta_1'=(\frac{\epsilon}{m})^{n-l}(1-\epsilon-\frac{\epsilon}{m})^{l-1}$, $\eta_2=(1-\epsilon-\frac{\epsilon}{m})^{n-1}$. When $a\notin\mathcal{L}$, according to Eq.\ref{us_final1}, we have
\begin{equation}
\begin{aligned}
\mathcal{U}^a(u^a_s,\tau^a) =& (1-\eta_1') (1-\alpha) Q(\bm{\acute{u}},\bm{\tau})-\sum_{k}^{m^{n-1}}\left[ \frac{p(u^a_s,u^{-a}_k)}{p(u^a_s)} \sum_{i}^{-a}\mathcal{U}^i(u^i_k,\tau^i)\right] +\eta_1'\mathcal{Q}_{its}(s, \bm{u}_s)\\
\label{us_general}
\end{aligned}
\end{equation}
As a result,
\begin{equation}
\mathcal{U}^a(u^a_s,\tau^a) = \left\{
\begin{array}{lr}
Eq.\ref{ugeneral} \ \ \ \ \ \ \ \ \ \ \ a\in \mathcal{L} \\ 
Eq.\ref{us_general} \ \ \ \ \ \ \ \ \ \ \ others
\end{array}
\right.\ \ \ \ \ \ \ \ \ \ 
\mathcal{U}^a(\acute{u}^a,\tau^a) = \left\{
\begin{array}{lr}
Eq.\ref{ugeneral} \ \ \ \ \ \ \ \ \ \ \ a\in \mathcal{L} \\ 
Eq.\ref{u*} \ \ \ \ \ \ \ \ \ \ \ others
\end{array}
\right.	\label{u_general}
\end{equation}
Therefore, 
\begin{equation}
\begin{aligned}
Q(\bm{u}_s,\bm{\tau})-Q(\bm{\acute{u}},\bm{\tau})&= \sum_a^{[1,n]-\mathcal{L}}\left[\mathcal{U}^a(u^a_s,\tau^a)-\mathcal{U}^a(\acute{u}^a,\tau^a)\right] \\
&=(n-l)(\eta_1'-\eta_2)\left[\mathcal{Q}(s, \bm{\acute{u}})-(1-\alpha)Q(\bm{\acute{u}},\bm{\tau})\right] + (n-l)\eta_1' e_Q\mathcal{Q}(s,\bm{\acute{u}})\label{delta_general}
\end{aligned}
\end{equation}

\section{STNs under ITS}
\subsection{The Optimal STN}
Suppose $\bm{\acute{u}}$ equals to the optimal action, i.e., $\bm{\acute{u}}=\bm{u}^*$. Notice the sum of coefficients of all true Q values in Eq.\ref{qfinalapp} equals
\begin{equation}
	2m\cdot\frac{\epsilon}{m}+2(1-\epsilon)-m^2\cdot\frac{\epsilon^2}{m^2}-2m\cdot\frac{\epsilon(1-\epsilon)}{m}-(1-\epsilon)^2 = 1\label{sum1}
\end{equation} 
In this case, there are only two kinds of actions: greedy action and inferior actions. Inspired by Eq.\ref{sum1}, $Q(\bm{\acute{u}},\bm{\tau})$ can be written as
\begin{equation}
	Q(\bm{\acute{u}},\bm{\tau})=w\mathcal{Q}(s,\bm{\acute{u}})+(1-w)\mathcal{Q}_{its}(s,\bm{u}_s)
\end{equation}
where $\mathcal{Q}_{its}(s,\bm{u}_s)=(1-\alpha)Q(\bm{\acute{u}},\bm{\tau})$ is the ITS target of inferior actions. Therefore, 
\begin{equation}
	\frac{\mathcal{Q}(s,\bm{\acute{u}})}{Q(\bm{\acute{u}},\bm{\tau})}=1+\alpha\frac{1-w}{w}\label{Qratio}
\end{equation}
We first consider the hardest exploration case. Substituting Eq.\ref{Qratio} into Eq.\ref{con_nagentsapp1}, we have
\begin{equation}
	\begin{aligned}
		Q(\bm{u}_s,\bm{\tau})-Q(\bm{\acute{u}},\bm{\tau})= n(\eta_1-\eta_2)\frac{\alpha}{w}Q(\bm{\acute{u}},\bm{\tau}) + n\eta_1 e_Q\mathcal{Q}(s,\bm{\acute{u}})\label{deltaQG}
	\end{aligned}
\end{equation}
Since $\bm{\acute{u}}=\bm{u}^*$, $\forall \bm{u}_s\neq \bm{\acute{u}}$, $\mathcal{Q}(s,\bm{u}_s)<\mathcal{Q}(s,\bm{\acute{u}})$ holds. Therefore, $e_Q<0$. Notice $\eta_1<\eta_2$, we have $Q(\bm{u}_s,\bm{\tau})-Q(\bm{\acute{u}},\bm{\tau})<0$, which indicates $Condition\ 1$ (Eq.5, Section 3.2) \textit{always holds under ITS}, i.e., ITS turn the optimal node into an STN. For non-hardest exploration case, the conclusion also holds.

\subsection{The Non-optimal STNs}
Suppose $\bm{\acute{u}}$ is a non-optimal action, $\exists \bm{u}_s$ such that $\mathcal{Q}(s,\bm{u}_s)>\mathcal{Q}(s,\bm{\acute{u}})$. According to $Conidtion\ 2$ (Eq.5, Section 3.2), to ensure this non-optimal point is not an STN, let $Q(\bm{u}_s,\bm{\tau})>Q(\bm{\acute{u}},\bm{\tau})$ and assume $Q(\bm{\acute{u}},\bm{\tau})\approx\mathcal{Q}(s, \bm{\acute{u}})$ (this assumption is quite accurate under ITS, as verified in Appendix F.2), for the hardest exploration case we have
\begin{equation}
\frac{\eta_1}{\eta_2}>\frac{\alpha}{\alpha+e_{Q0}}\label{e_00}
\end{equation}
where $e_{Q0}\in[0,\infty)$ is a hyper-parameter that defines the minimum gap of the true Q values between the superior and greedy actions. Therefore, \textit{the non-optimal STNs can be eliminated by raising $\frac{\eta_1}{\eta_2}$ under ITS}.

For the non-hardest exploration case, $\eta_1$ in Eq.\ref{e_00} is replaced by $\eta_1'$. Since $\frac{\eta_1}{\eta_2}>\frac{\eta_1'}{\eta_2}$, we only need to consider the hardest exploration case.

\section{STNs under ITS with Superior Sample Weight}
\subsection{Derivation of $Q(\bm{u}_s,\bm{\tau})-Q(\bm{\acute{u}},\bm{\tau})$}
Given the greedy action $\{\acute{u}^1,\cdots,\acute{u}^n\}=\bm{\acute{u}}$ and a superior action $\{u^1_s,\cdots,u^n_s\}=\bm{u}_s\neq \bm{\acute{u}}$, we have $\mathcal{Q}(s,\bm{u}_s)>\mathcal{Q}(s,\bm{\acute{u}})$. Here we regard all actions except $\bm{u}_s$ and $\bm{\acute{u}}$ as inferior actions. Discussions under multiple superior actions is provided in Appendix H.1. We first consider the hardest exploration case. For simplicity, assuming $Q(\bm{\acute{u}},\bm{\tau})>0$, where $\mathcal{Q}_{its}(s, \bm{u})=(1-\alpha)Q(\bm{\acute{u}},\bm{\tau})$ for inferior actions. By applying a weight $w$ on the superior action, the utility function of the individual action $u^a_s(a\in[1,n])$ consists of two parts
\begin{equation}
\begin{aligned}
\mathcal{U}^a(u^a_s,\tau^a) =& (1-\eta_{1,w})\left[ (1-\alpha) Q(\bm{\acute{u}},\bm{\tau})-\sum_{k}^{m^{n-1}-1}\left[ \frac{p(u^a_s,u^{-a}_k)}{p(u^a_s)-p(\bm{u}_s)} \sum_{i}^{-a}\mathcal{U}^i(u^i_k,\tau^i)\right] \right]\\
&+\eta_{1,w}\left[\mathcal{Q}_{its}(s, \bm{u}_s)-\sum_i^{-a} \mathcal{U}^i(u^i_s,\tau^i) \right]\\
\label{usw}
\end{aligned}
\end{equation}
where $\eta_{1,w}=\frac{w\eta_1}{1+(w-1)\eta_1}$, $\eta_1=(\frac{\epsilon}{m})^{n-1}$. Please refer to Appendix D.2 for more details about the notations. According to Eq.\ref{ee1}, we have
\begin{equation}
\begin{aligned}
\sum_{k}^{m^{n-1}-1} \left[\frac{p(u^a_s,u^{-a}_k)}{p(u^a_s)-p(\bm{u}_s)} \sum_{i}^{-a}\mathcal{U}^i(u^i_k,\tau^i)\right] = \frac{1}{(1-\eta_1)}\sum_{k}^{m^{n-1}} \left[\frac{p(u^a_s,u^{-a}_k)}{p(u^a_s)} \sum_{i}^{-a}\mathcal{U}^i(u^i_k,\tau^i)\right] - \frac{\eta_1}{(1-\eta_1)}\sum_{i}^{-a}\mathcal{U}^i(u^i_s,\tau^i) \\
\label{eew}
\end{aligned}
\end{equation}
Substituting Eq.\ref{eew} into Eq.\ref{usw}, we have
\begin{equation}
\begin{aligned}
\mathcal{U}^a(u^a_s,\tau^a) =& (1-\eta_{1,w}) (1-\alpha) Q(\bm{\acute{u}},\bm{\tau})-\frac{1-\eta_{1,w}}{1-\eta_1}\sum_{k}^{m^{n-1}}\left[ \frac{p(u^a_s,u^{-a}_k)}{p(u^a_s)} \sum_{i}^{-a}\mathcal{U}^i(u^i_k,\tau^i)\right]\\&+
 \frac{\eta_1-\eta_{1,w}}{1-\eta_1}\sum_{i}^{-a}\mathcal{U}^i(u^i_s,\tau^i) +\eta_{1,w}\mathcal{Q}_{its}(s, \bm{u}_s)
\label{us_final1w}
\end{aligned}
\end{equation}
Notice that
\begin{equation}
\begin{aligned}
\sum_{a=1}^{n}\sum_{i}^{-a}\mathcal{U}^i(u^i_s,\tau^i)=(n-1)\sum_{a=1}^{n}\mathcal{U}^i(u^i_s,\tau^i)=(n-1)Q(\bm{u}_s,\bm{\tau})
\label{Qsw}
\end{aligned}
\end{equation}
Therefore,
\begin{equation}
\begin{aligned}
Q(\bm{u}_s,\bm{\tau}) = \sum_{a=1}^n \mathcal{U}^a(u^a_s,\tau^a) =& n(1-\eta_{1,w}) (1-\alpha) Q(\bm{\acute{u}},\bm{\tau}) +(n-1)\frac{\eta_1-\eta_{1,w}}{1-\eta_1}Q(\bm{u}_s,\bm{\tau})\\
&+n\eta_{1,w}\mathcal{Q}_{its}(s, \bm{u}_s)-\frac{1-\eta_{1,w}}{1-\eta_1} \sum_{a=1}^n \sum_{k}^{m^{n-1}} \left[\frac{p(u^a_s,u^{-a}_k)}{p(u^a_s)} \sum_{i}^{-a}\mathcal{U}^i(u^i_k,\tau^i)\right] 
\label{Qsw}
\end{aligned}
\end{equation}
According to Eq.\ref{Q*1}, we have
\begin{equation}
\begin{aligned}
\sum_{a=1}^n \sum_{k}^{m^{n-1}} \left[\frac{p(\acute{u}^a,u^{-a}_k)}{p(\acute{u}^a)} \sum_{i}^{-a}\mathcal{U}^i(u^i_k,\tau^i)\right]&= n(1-\eta_2) (1-\alpha) Q(\bm{\acute{u}},\bm{\tau}) +n\eta_2\mathcal{Q}(s, \bm{\acute{u}})-Q(\bm{\acute{u}},\bm{\tau}) \label{eew2}
\end{aligned}
\end{equation}
Substituting Eq.\ref{eew2} into Eq.\ref{Qsw}, we have
\begin{equation}
\begin{aligned}
\left[1-(n-1)\frac{\eta_1-\eta_{1,w}}{1-\eta_1}\right] Q(\bm{u}_s,\bm{\tau}) =& n(1-\eta_{1,w}) (1-\alpha) Q(\bm{\acute{u}},\bm{\tau})+n\eta_{1,w}\mathcal{Q}_{its}(s, \bm{u}_s)\\
&-\frac{1-\eta_{1,w}}{1-\eta_1} \left[n(1-\eta_2) (1-\alpha)-1\right] Q(\bm{\acute{u}},\bm{\tau}) - n\eta_2\frac{1-\eta_{1,w}}{1-\eta_1} \mathcal{Q}(s, \bm{\acute{u}}) 
\label{Qsw2_0}
\end{aligned}
\end{equation}
where $\eta_2=(1-\epsilon+\frac{\epsilon}{m})^{n-1}$. Eq.\ref{Qsw2_0} can be further simplified
\begin{equation}
\begin{aligned}
Q(\bm{u}_s,\bm{\tau}) =& \frac{n(1-\alpha)(\eta_2-\eta_1)+1}{1+n(w-1)\eta_1}Q(\bm{\acute{u}},\bm{\tau})
+n\frac{w(1+e_Q)\eta_1-\eta_2}{1+n(w-1)\eta_1} \mathcal{Q}(s, \bm{\acute{u}}) 
\label{Qsw2}
\end{aligned}
\end{equation}
Finally, we have
\begin{equation}
\begin{aligned}
Q(\bm{u}_s,\bm{\tau})-Q(\bm{\acute{u}},\bm{\tau})= n\frac{(1-\alpha)(\eta_2-\eta_1)-(w-1)\eta_1}{1+n(w-1)\eta_1}Q(\bm{\acute{u}},\bm{\tau})+n\frac{w(1+e_Q)\eta_1-\eta_2}{1+n(w-1)\eta_1} \mathcal{Q}(s, \bm{\acute{u}})\label{deltaQwfinal}
\end{aligned}
\end{equation}
\textit{When $\bm{w=1}$, Eq.\ref{deltaQwfinal} degenerates to Eq.\ref{con_nagentsapp1}}. For monotonic value decomposition, Eq.\ref{deltaQwfinal} also holds since $Q(\bm{\acute{u}}, \bm{\tau})$ and $Q(\bm{u}_s, \bm{\tau})$ do not change. 

Since $\bm{\acute{u}}$ is a non-optimal action, according to $Conidtion\ 2$ (Eq.5, Section 3.2), to ensure this non-optimal point is not an STN, let $Q(\bm{u}_s,\bm{\tau})>Q(\bm{\acute{u}},\bm{\tau})$ and assume $Q(\bm{\acute{u}},\bm{\tau})\approx\mathcal{Q}(s, \bm{\acute{u}})$ (this assumption is quite accurate under ITS, as verified in Appendix F.2), we have
\begin{equation}
w>\frac{\alpha(\eta_2-\eta_1)}{e_{Q0}\eta_1}=w_0\label{w_0}
\end{equation}
Therefore, the non-optimal STNs can be eliminated by applying a large enough weight on the superior actions under ITS. For the non-hardest exploration cases, $\eta_1$ in Eq.\ref{w_0} is replaced with $\eta_1'$. Since  $\frac{\alpha(\eta_2-\eta_1)}{e_{Q0}\eta_1}>\frac{\alpha(\eta_2-\eta_1')}{e_{Q0}\eta_1'}$, we only need to consider the hardest exploration case.

\subsection{Verification}
We carry out experiments in matrix games to evaluate the effect of weights on the superior actions under ITS. The pay-off matrix is defined as
\begin{equation}
\mathcal{Q}(s,\bm{u})=
\left\{
\begin{array}{lr}
6(1+e_Q) \ \ \ \ \ \ \ \ \ \ \ \ \ \ \ \ \bm{u}=\{0,0\}\\ 
6 \ \ \ \ \ \ \ \ \ \ \ \ \ \ \ \ \ \ \ \ \ \ \ \ \ \ \ \ \ \ \ \bm{u}=\{m,m\} \\
random(-20, 6) \ \ \ \ \ others
\end{array}
\right.	\label{weight_mat}
\end{equation}
where $m$ is the size of individual action space. The greedy action is fixed to $\bm{\acute{u}}=\{m,m\}$. Therefore,  $Q(\bm{\acute{u}},\bm{\tau})=6$. An mlp shared by all agents is adopted as the agent network. 1000 iterations (100 episodes per iteration) of training over 5 seeds are executed on each set of parameters, where $\alpha=0.1$, $\epsilon=0.2$ and $e_Q=1/3$. 

\begin{minipage}{1.0\linewidth}
\tabcaption{Comparison between test and calculation (shown in parentheses) $Q(\bm{u}_s,\bm{\tau})-Q(\bm{\acute{u}},\bm{\tau})$ on $n$-agent matrix games when $w=w_0$.}\label{tabxx}
\vskip 0.15in
\begin{center}
\begin{small}
\begin{tabular}{c|ccccc}
 \toprule
$m ^ n$ & $3^2$ & $5^2$ & $10^2$ & $3^3$ & $3^4$\\
 \midrule
     $w_0$ (Eq.\ref{w_0}) & 3.60 & 6.00 & 12.00 & 50.32 & \textbf{659.50} \\ 
		   $Q(\bm{u}_s,\bm{\tau})-Q(\bm{\acute{u}},\bm{\tau})$ (Eq.\ref{deltaQwfinal}) & \makecell[c]{0.01 $\pm$0.06\\(0)} & \makecell[c]{0.02 $\pm$0.16\\(0)} & \makecell[c]{0.22 $\pm$0.13\\(0)} & \makecell[c]{-0.02 $\pm$0.30\\(0)} & \makecell[c]{-0.48 $\pm$0.75\\(0)}\\
            Test $Q(\bm{\acute{u}},\bm{\tau})$ & 5.95 $\pm$0.02& 5.97$\pm$0.02& 5.98 $\pm$0.01& 5.90 $\pm$0.06& 5.93 $\pm$0.03\\
            \bottomrule
        \end{tabular}
\end{small}
\end{center}
\vskip -0.2in
\end{minipage}

The experimental results are shown in Tab.\ref{tabxx}. Firstly, the error of $Q(\bm{u}_s,\bm{\tau})-Q(\bm{\acute{u}},\bm{\tau})$ between test and calculation is very small. Secondly, the joint Q value of the greedy action approximates its true Q value, i.e., $Q(\bm{\acute{u}},\bm{\tau})\approx\mathcal{Q}(s,\bm{\acute{u}})=6$) under ITS. Thirdly, the lower bound of $w$ grows \textit{exponentially} as the number of agent grows, which introduces instability in $Q(\bm{u}_s,\bm{\tau})-Q(\bm{\acute{u}},\bm{\tau})$.

\section{STNs under ITS with Superior Experience Replay}
Given the greedy action $\{\acute{u}^1,\cdots,\acute{u}^n\}=\bm{\acute{u}}$ and a superior action $\{u^1_s,\cdots,u^n_s\}=\bm{u}_s\neq \bm{\acute{u}}$, we have $\mathcal{Q}(s,\bm{u}_s)>\mathcal{Q}(s,\bm{\acute{u}})$. Here we regard all actions except $\bm{u}_s$ and $\bm{\acute{u}}$ as inferior actions. Discussions under multiple superior actions is provided in Appendix H.1. We first consider the hardest exploration case. For simplicity, assuming $Q(\bm{\acute{u}},\bm{\tau})>0$, where $\mathcal{Q}_{its}(s, \bm{u})=(1-\alpha)Q(\bm{\acute{u}},\bm{\tau})$ for inferior actions. By applying a weight $w_{ser}$ on the loss of superior actions from the superior buffer, the utility function of individual action $u^a_s(a\in[1,n])$ consists of two parts
\begin{equation}
\begin{aligned}
\mathcal{U}^a(u^a_s,\tau^a) =& (1-\eta_{1,ser})\left[ (1-\alpha) Q(\bm{\acute{u}},\bm{\tau})-\sum_{k}^{m^{n-1}-1}\left[ \frac{p(u^a_s,u^{-a}_k)}{p(u^a_s)-p(\bm{u}_s)} \sum_{i}^{-a}\mathcal{U}^i(u^i_k,\tau^i)\right] \right]\\
&+\eta_{1,ser}\left[\mathcal{Q}_{its}(s, \bm{u}_s)-\sum_i^{-a} \mathcal{U}^i(u^i_s,\tau^i) \right]
\label{usser}
\end{aligned}
\end{equation}
where $\eta_{1,ser}=\frac{w_{ser}+\eta_1\eta_s}{\eta_s+w_{ser}}$. $\eta_1=(\frac{\epsilon}{m})^{n-1}$, and $\eta_s$ is the probability of state $s$. Please refer to Appendix D.2 for more details about the notations. Following the derivation provided in Appendix F.1, we have
\begin{equation}
\begin{aligned}
\left[1-(n-1)\frac{\eta_1-\eta_{1,ser}}{1-\eta_1}\right] Q(\bm{u}_s,\bm{\tau})=& n(1-\eta_{1,ser}) (1-\alpha) Q(\bm{\acute{u}},\bm{\tau})+n\eta_{1,ser}\mathcal{Q}_{its}(s, \bm{u}_s)\\
&-\frac{1-\eta_{1,ser}}{1-\eta_1} \left[n(1-\eta_2) (1-\alpha)-1\right] Q(\bm{\acute{u}},\bm{\tau}) - n\eta_2\frac{1-\eta_{1,ser}}{1-\eta_1} \mathcal{Q}(s, \bm{\acute{u}}) 
\label{Qswser}
\end{aligned}
\end{equation}
Eq.\ref{Qswser} can be further simplified
\begin{equation}
\begin{aligned}
Q(\bm{u}_s,\bm{\tau}) = \eta_s\frac{n(1-\alpha)(\eta_2-\eta_1)+1}{\eta_s+nw_{ser}} Q(\bm{\acute{u}},\bm{\tau})+n\frac{(w_{ser}+\eta_1\eta_s)(1+e_Q)-\eta_2\eta_s}{\eta_s+nw_{ser}} \mathcal{Q}(s, \bm{\acute{u}}) 
\label{Qsfinalser}
\end{aligned}
\end{equation}
where $\eta_2=(1-\epsilon+\frac{\epsilon}{m})^{n-1}$. Therefore, we have
\begin{equation}
\begin{aligned}
Q(\bm{u}_s,\bm{\tau})-Q(\bm{\acute{u}},\bm{\tau})= n\frac{(1-\alpha)(\eta_2-\eta_1)\eta_s-w_{ser}}{\eta_s+nw_{ser}}Q(\bm{\acute{u}},\bm{\tau})+n\frac{(w_{ser}+\eta_1\eta_s)(1+e_Q)-\eta_2\eta_s}{\eta_s+nw_{ser}} \mathcal{Q}(s, \bm{\acute{u}}) \label{deltaQser}
\end{aligned}
\end{equation}
\textit{When $\bm{w=0}$, Eq.\ref{deltaQser} degenerates to Eq.\ref{con_nagentsapp1}}. For monotonic value decomposition, Eq.\ref{deltaQser} also holds since $Q(\bm{\acute{u}}, \bm{\tau})$ and $Q(\bm{u}_s, \bm{\tau})$ do not change. 

Since $\bm{\acute{u}}$ is a non-optimal action, according to $Conidtion\ 2$ (Eq.5, Section 3.2), to ensure this non-optimal point is not an STN, let $Q(\bm{u}_s,\bm{\tau})>Q(\bm{\acute{u}},\bm{\tau})$ and assume $Q(\bm{\acute{u}},\bm{\tau})\approx\mathcal{Q}(s, \bm{\acute{u}})$ (this assumption is quite accurate under ITS, as verified in Appendix F.2), we have
\begin{equation}
w_{ser}>\frac{\alpha}{e_{Q0}}(\eta_2-\eta_1)\eta_s-\eta_1\eta_s\label{conditionserapp}
\end{equation}
According to Eq.\ref{conditionserapp}, SER can eliminate the non-optimal STNs by selecting a large enough $\bm{w}$ for the loss of superior actions from the superior buffer. For the non-hardest exploration cases, $\eta_1$ in Eq.\ref{w_0} is replaced with $\eta_1'$. Since  $\eta_1'>\eta_1$, we only need to consider the hardest exploration case.

\section{Discussions}
\subsection{GVR under Multiple Superior Actions}
\begin{figure}[h]
	\centering
	\includegraphics[scale=0.16]{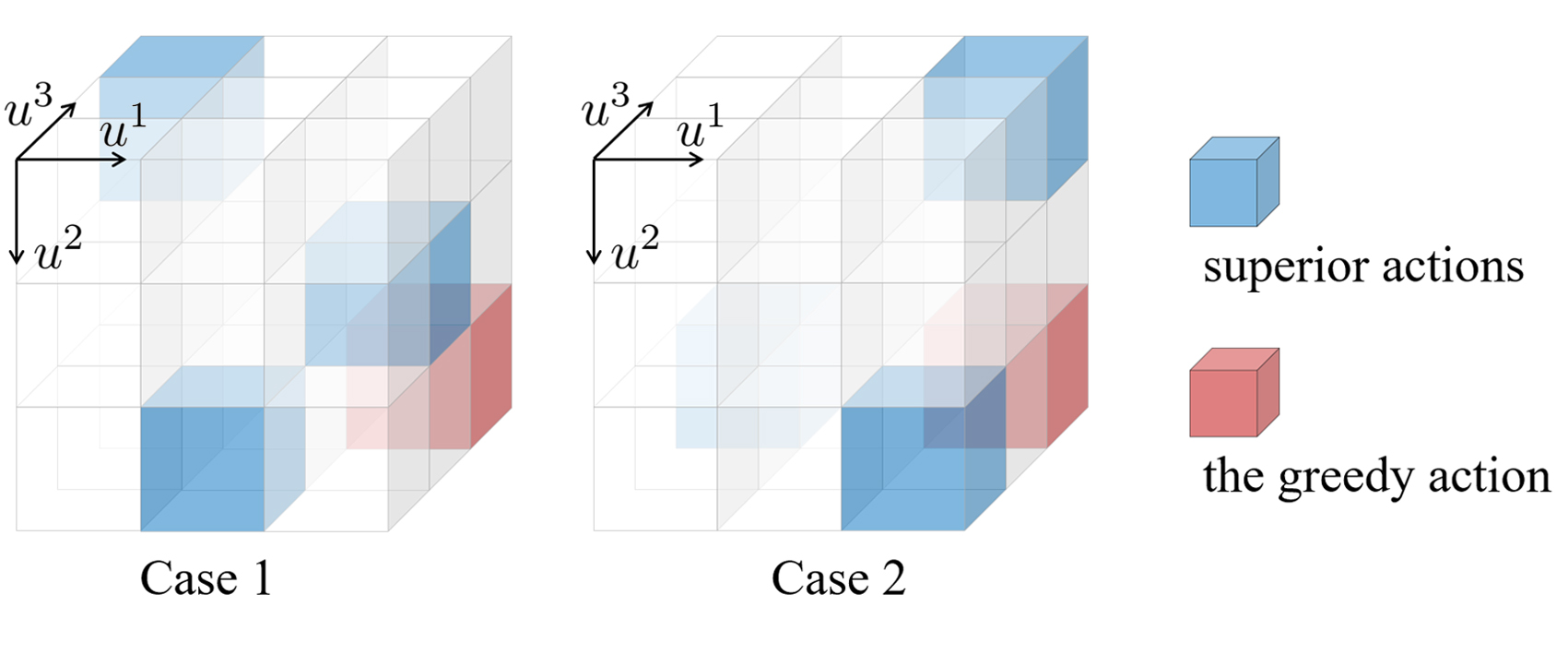}
	\vspace{-0.1cm}
	\caption{Failure cases for condition in Eq.\ref{e_00} under multiple superior actions, where the number of agents and the size of individual action space are both 3. The superior actions ($\{\bm{u}_{s,1},\cdots,\bm{u}_{s,p}\}$) and greedy action $\bm{\acute{u}}$) are denoted by blocks of blue and red respectively.}\label{morethanone}
\end{figure}

In previous derivations, we regard all actions except $\bm{u}_s$ and $\bm{\acute{u}}$ as inferior actions, i.e., we only consider the cases with no more than one superior action. For situations with multiple superior actions $\{\bm{u}_{s,1},\cdots,\bm{u}_{s,p}\}$, two examples where the condition in Eq.\ref{e_00} fails to eliminate the non-optimal STNs are given in Fig.\ref{morethanone}.

In both examples, $Q(\bm{u}_s,\bm{\tau})-Q(\bm{\acute{u}},\bm{\tau})$ is smaller than any cases with only one superior action. Eq.\ref{e_00} is an insufficient condition for $Q(\bm{u}_s,\bm{\tau})>Q(\bm{\acute{u}},\bm{\tau})$. Since most of the superior actions in training batch come from superior buffer, to avoid the cases with multiple superior actions, we set the superior batch size to 1. 

\subsection{Trade-off between Optimality and Stability}
\begin{figure}[h]
	\centering
	\includegraphics[width=0.7\columnwidth]{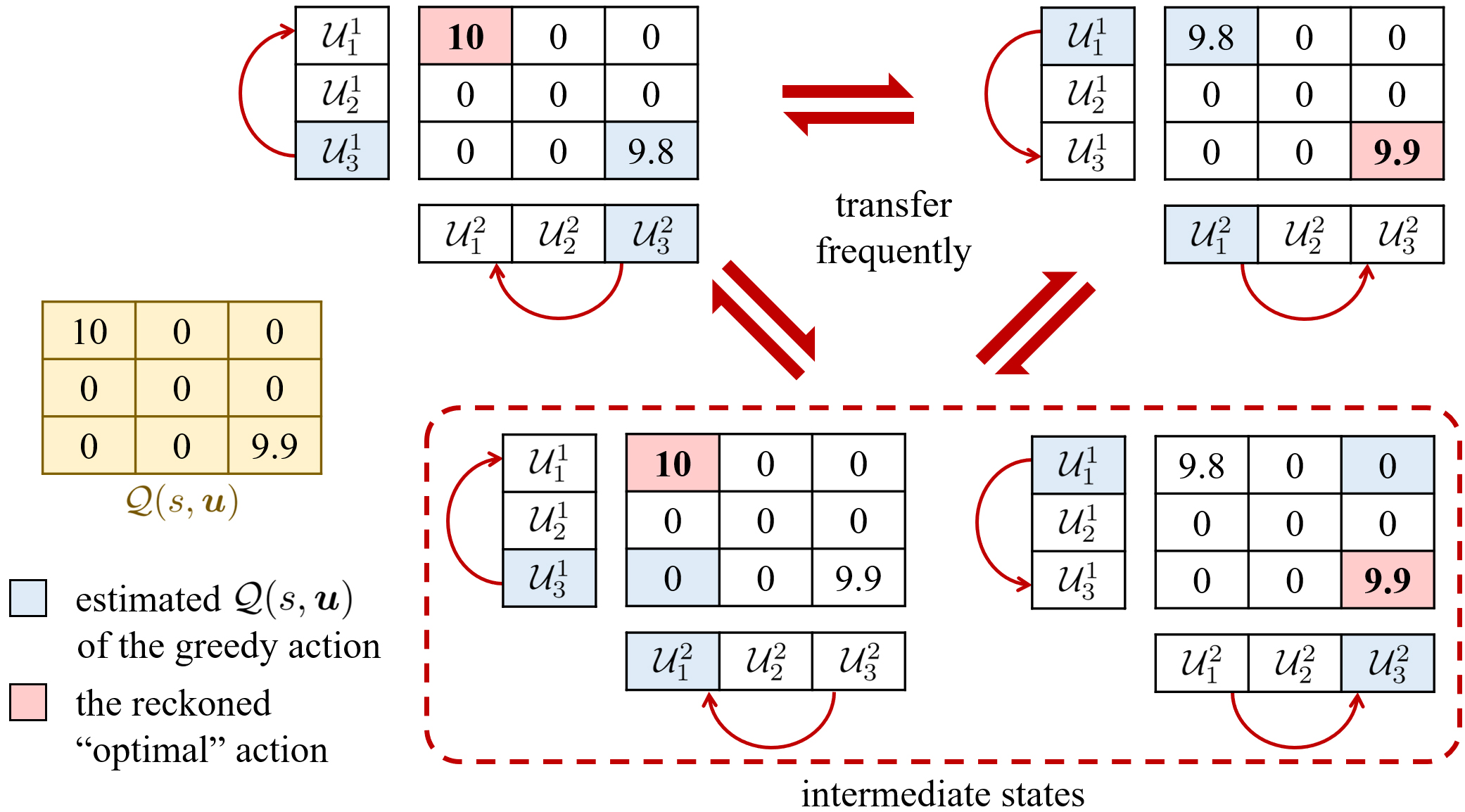}
	\vspace{-0.1cm}
	\caption{An example of instability caused by excessive pursuit for optimality. Due to the estimation error of the true Q values, actions with large true Q values may be mistaken for the optimal action. As a result, the joint policy transfers frequently between the reckoned "optimal" actions. Worse, the out-of-step updates of individual polices put the joint policy at poor intermediate states.}\label{tradeoffexam}
\end{figure}

The optimal consistency (or TGM principle) requires access to the true Q values, which are usually obtained by estimation. In such cases, excessive pursuit for optimality may decrease the stability. An example is shown in Fig.\ref{tradeoffexam}.

\section{The Working Principle of GVR}
\begin{figure}[h]
	\centering
	\includegraphics[width=0.65\columnwidth]{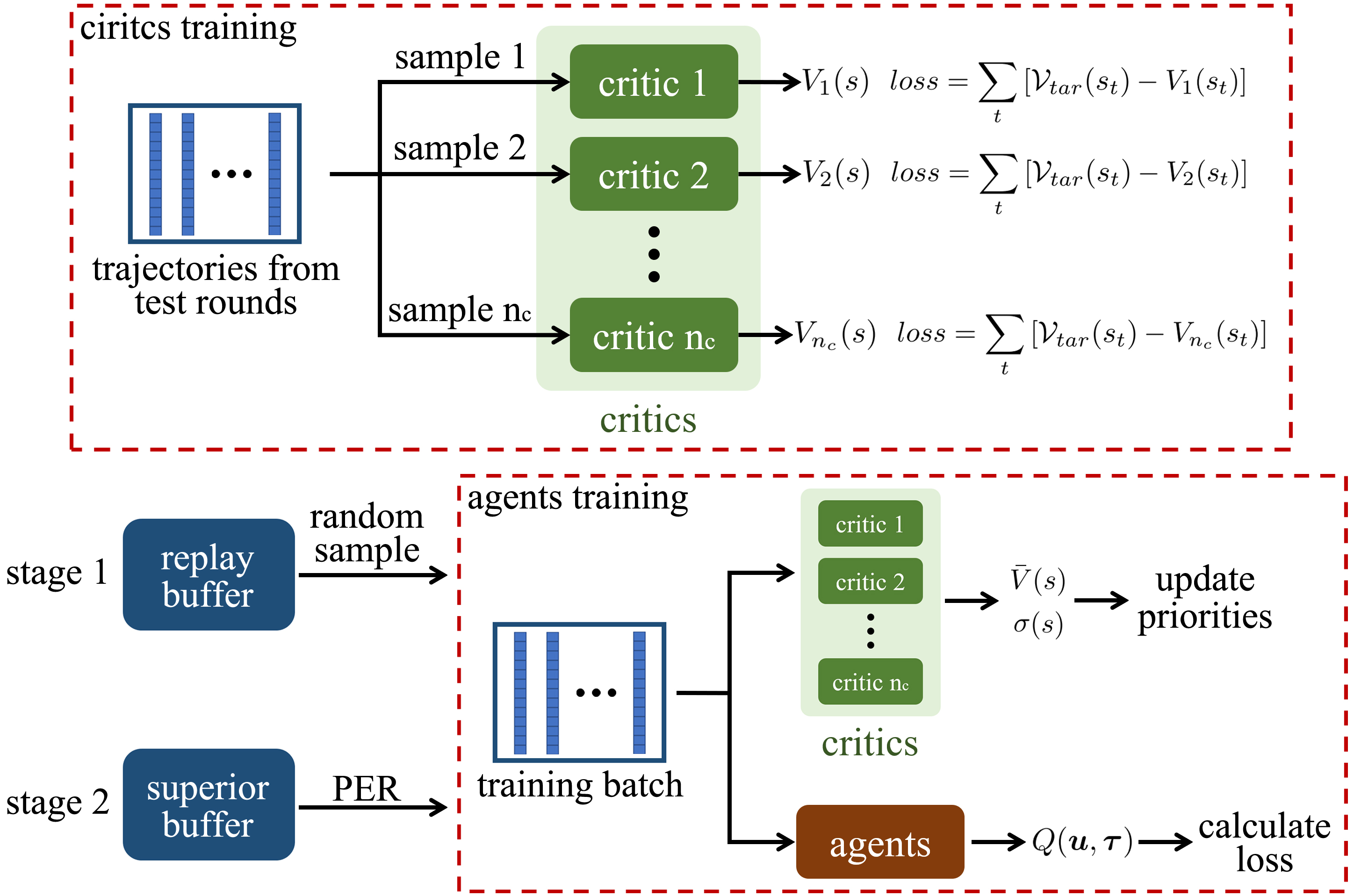}
	\caption{The working principle of GVR. The critics are trained on the trajectories sampled from test rounds. All critics share the same target $\mathcal{V}_{gvr}(s)$ (Eq.11). The agents training consists of two stages. In stage 1, the training batch is randomly sampled from the replay buffer, where the loss function is $loss = \sum_t \left[ \mathcal{Q}_{its}(s_t,\bm{u}_t)-Q({\bm{u}_t,\bm{\tau}_t})\right]$. In stage 2, the training batch is sampled from the superior buffer with prioritized experience replay (PER), where the loss function is $loss = \sum_t w(s_t)\mathbbm{I}_{sup}\left[\mathcal{Q}_{its}(s_t,\bm{u}_t)- Q(\bm{u}_t,\bm{\tau}_t) \right]$. $\mathbbm{I}_{sup}(s_t,\bm{u}_t)$ is a indicator for the superior action, i.e., $\mathbbm{I}_{sup}(s_t,\bm{u}_t)=1$, $s.t. \sum_{t=t_0}^T \gamma^{t-t_0} r(s_t,\bm{u}_t)>\bar{V}(s_t)+3\sigma(s_t)$. $w_{ser}(s_t)=\frac{\alpha}{e_{Q0}}(\eta_2-\eta_1)-\eta_1$ for hardest exploration cases and $w_{ser}(s_t)=\frac{\alpha}{e_{Q0}}(\eta_2-\eta_1')-\eta_1'$ for non-hardest exploration cases, where $e_{Q0}=\frac{3\sigma(s)}{\bar{V}(s)}$. At the end of both stages, the trajectories in the training batch are stored into the superior buffer after the update of their priorities.}
	\label{process}
\end{figure}

The working principle of GVR is shown in Fig.\ref{process}. Please refer to Appendix G for more details about the notations. 

\section{Experimental Settings and Additional Experiments}
\subsection{Experimental Settings}
In experiments on one-step matrix games, since the episode length is $1$, an mlp shared by all agents is adopted as the agent network. Besides, $\eta_s=1$ since there is only 1 state. No data buffer is used in the verification of calculation results, e.g., $Q(\bm{u}_s,\bm{\tau})-Q(\bm{\acute{u}},\bm{\tau})$ (Fig.2(a), Section 5.1) and $Q(\bm{u},\bm{\tau})$ (Tab.\ref{tab2}, Appendix C.2). Otherwise, a replay buffer of length 1000 of is applied matrix for all algorithms, e.g., Fig.2(b) (Section 5.1) and Fig.3 (Section 5.1). For WQMIX, $\alpha=0.5$. For GVR, $\alpha=0.2$ and the length of superior buffer is 3. All experiments are carried out over 5 seeds.

In experiments on predator-prey and SMAC, we adopt the default settings for VDN, QMIX, QPLEX and WQMIX. The length of replay buffer is 5000 and the batch size is 32. For WQMIX, $\alpha=0.5$. For GVR, $\alpha=0.2$ and the length of superior buffer is 300. According to Appendix H.1, we set the size of superior batch to 1. The probability of a state $\eta_s$ is unknown, a sufficient condition of Eq.\ref{conditionserapp} is $\eta_s=1$. All experiments are carried out over 5 seeds.

The game version of StarCraft II is 69232. Each algorithm is trained for 2e6 steps in MMM2, 2c\_vs\_64\_zg and 6h\_vs\_8z, with $\epsilon$ damping from 1 to 0.05 in the first 5e4 steps. Besides, in 6h\_vs\_8z, 3s5z\_vs\_3s6z and corridor, each algorithm is trained for 5e6 steps, with $\epsilon$ damping from 1 to 0.05 in the first 1e6 steps.

\begin{figure}[t]
    \vskip 0.2in
    \begin{center}
    \centerline{\includegraphics[width=\columnwidth]{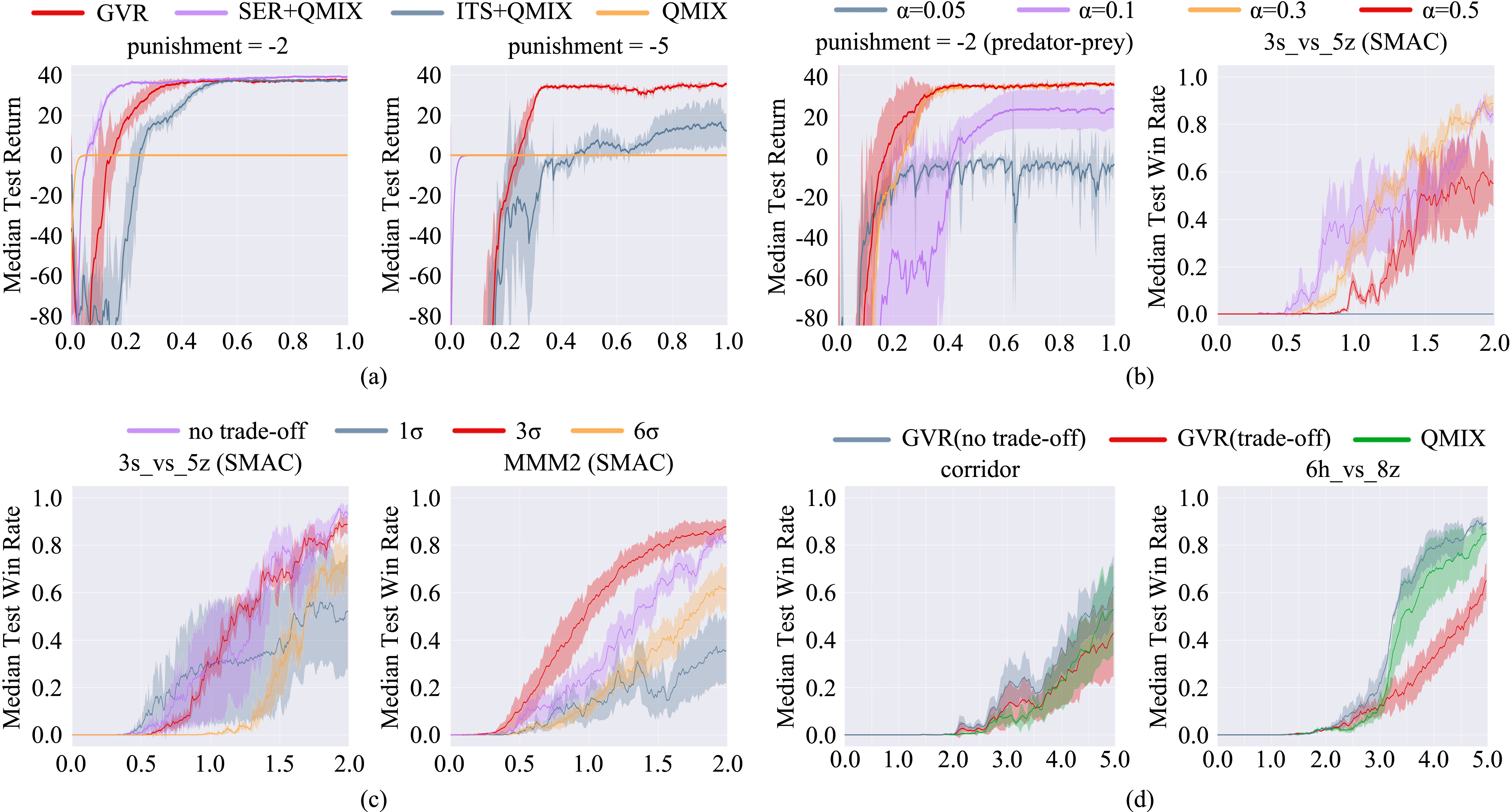}}
    \caption{Ablation studies. The x-axes are training time-steps (million). (a) Effect of ITS and SER. Investigation on parameter (b) $\alpha$ and (c) $e_{Q0}$. (d) GVR with trade-off ($3\sigma$) vs GVR without trade-off in hard exploration tasks.}\label{ab}
    \end{center}
    \vskip -0.2in
\end{figure}

\subsection{Ablation Studies}
We conduct ablation studies to investigate the improvements of GVR. We first evaluate the effect of inferior target shaping (ITS) and superior experience replay (SER) on predator-prey. We apply a constant weight $w=3$ to the loss of action $\bm{u_t}$ for ITS when $Q(\bm{u_t},\bm{\tau}_t)<r(s_t,\bm{u_t})+\gamma Q(\bm{u_{t+1}},\bm{\tau}_{t+1})$. As shown in Fig.\ref{ab}(a), in task with punishment -2, both ITS and SER help to overcome the relative overgeneralization (RO). In task with punishment -5, SER alone is unable to overcome the RO since the STNs depend on the true Q values of inferior actions. Although ITS remove the dependence, it can not ensure the optimal consistency without SER.

\begin{figure}[hbp]
	\vskip 0.2in
    \begin{center}
    \centerline{\includegraphics[width=\columnwidth]{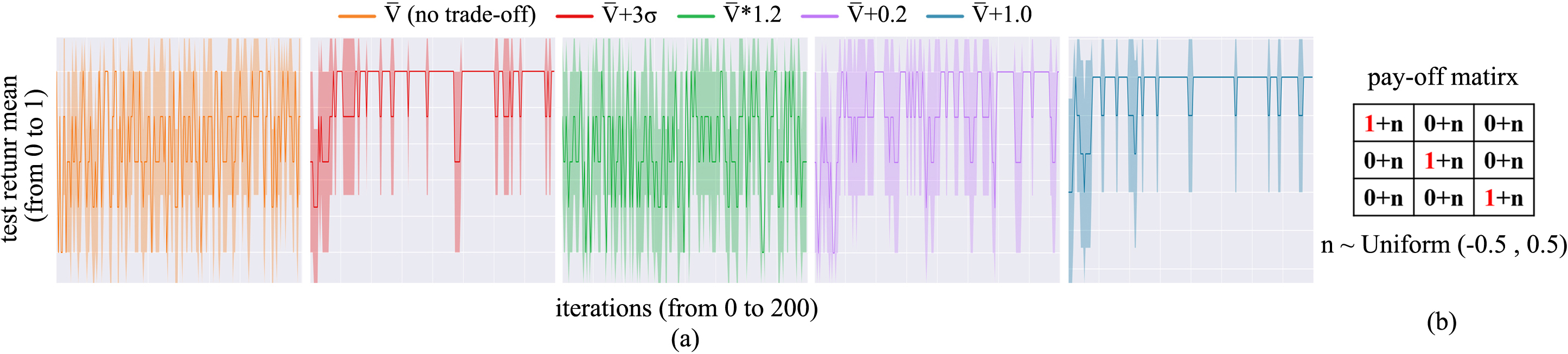}}  
    \caption{Investigations of threshold functions in a two-agent matrix game. The optimal node changes in training due to a noise applied to the rewards. As a result, the greedy action may jump frequently between the reckoned optimal nodes and poor intermediates.}
    \end{center}
    \vskip -0.2in
\end{figure}

We also investigate the effect of the parameter $\alpha$. As shown in Fig.\ref{ab}(b), a too small or too large $\alpha$ leads to poor performance. Since $\alpha$ defines the gap of the joint Q values' targets between the inferior and greedy actions, a too small $\alpha$ causes confusion between the greedy and the inferior actions. Meanwhile, a too large $\alpha$ prevents the update from a greedy action to a superior action.

To find a suitable hyper-parameter for the trade-off between stability and optimality, we compare the performance of GVR under different $e_{Q0)}$. The experimental results are shown in Fig.\ref{ab}(c), where $k\sigma$ denotes $e_{Q0}=\frac{k\sigma(s)}{\bar{V}(s)}$ ($k\in\{1,3,6\}$). The algorithm attaches more importance to stability as $k$ increases. In the experiments of $no\ trade-off$, we do not use the ensemble critics, where the joint Q value function is applied to identify inferior and superior actions. In this case, an action $\bm{u}_t$ is classified into superior action when $\bm{u}_t\neq\bm{\acute{u}}$ and $r(s_t,\bm{u_t})+\gamma Q(\bm{u_{t+1}},\bm{\tau}_{t+1})>Q(\bm{\acute{u}_t},\bm{\tau}_t)$. As shown in Fig.\ref{ab}(c), a proper $e_{Q0}$ is helpful for the balance of stability and optimality. However, the trade-off do not always helps. As shown in Fig.\ref{ab}(d), in hard exploration tasks, GVR with trade-off do not perform better than GVR without trade-off. 

In the experiments of GVR with trade-off, we find the critics often conflict with the joint Q value function, e.g., $r(s_t, \bm{u}_t)+\gamma Q(\bm{u}_{t+1},\bm{\tau}_{t+1})<Q(\bm{u}_t,\bm{\tau}_t)$ happens for a superior action $\bm{u}_t$. The ensemble critics and the joint Q value function are both evaluations of the joint policy. We mix these two evaluations to the achieve the trade-off in GVR. The conflict between two evaluations may cause performance deterioration. As a result, GVR with trade-off performs worse than that without trade-off in some tasks. The latter adopts the joint Q value function as the unique evaluation.

For the adaptive trade-off, we investigate some other threshold functions, such as $\bar{V}+C$ and $(1+C) * \bar{V}$, where $C>0$ is a predefined parameter. However, the scalability of these threshold functions is poor because (1) the suitable $C$ varies as the reward scale; (2) the estimation error of the critics decreases as training, where $C$ requires attenuation accordingly. Empirical results are shown in Fig.1. A well-designed threshold ($\bar{V}+1.0$) may perform better but requires prior knowledge about the value's gap between the optimal and sub-optimal actions.

\subsection{Comparison With Joint Exploration Methods}
\begin{figure}[h]
    \vskip 0.2in
    \begin{center}
    \centerline{\includegraphics[width=\columnwidth]{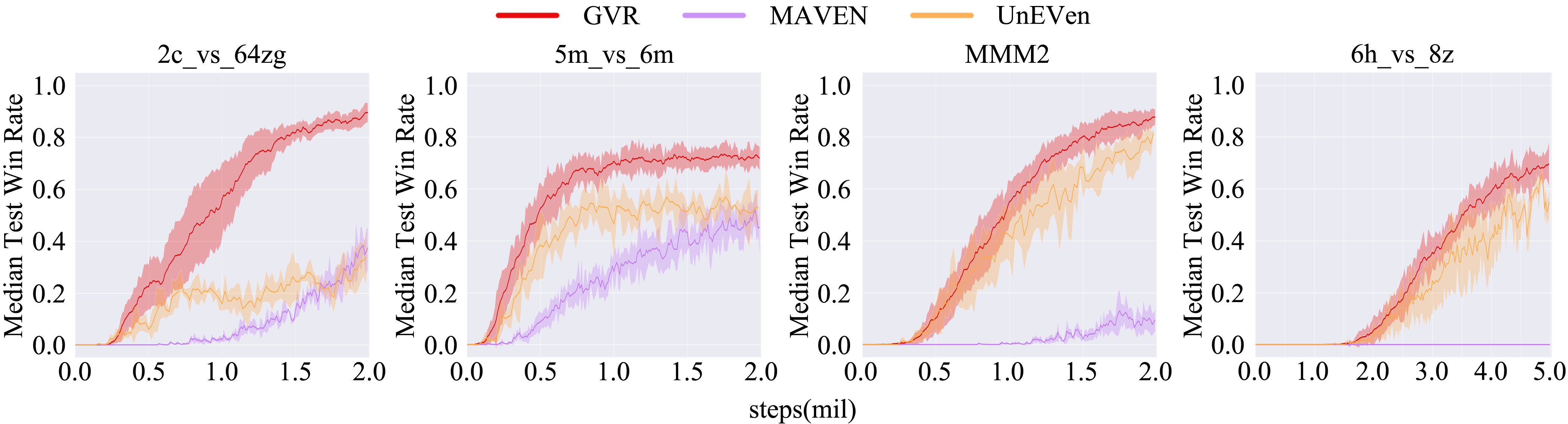}}
    \caption{Comparison between GVR, MAVEN and UneVEn.}\label{smac2}
    \end{center}
    \vskip -0.2in
\end{figure}
We compare our method with joint exploration methods on SMAC. The experimental results are shown in Fig.\ref{smac2}.


\end{document}